\documentclass[preprint,12pt]{elsarticle}
\usepackage{amsmath,amssymb}
\usepackage{amsthm}
\usepackage{mathtools}
\usepackage{isomath}
\usepackage{mathrsfs}

\usepackage{graphicx} 
\usepackage{caption}
\usepackage{subcaption}
\usepackage{hyperref}
\usepackage[dvipsnames]{xcolor}
\usepackage{siunitx}
\usepackage{algorithm}
\usepackage{algpseudocode}
\usepackage{changes}

\usepackage{multirow}

\newcommand{\vecx}{\mathbfit{x}}
\newcommand{\vecy}{\mathbfit{y}}
\newcommand{\vecr}{\mathbfit{r}}
\newcommand{\pcx}{\mathbfit{X}}
\newcommand{\pcy}{\mathbfit{Y}}

\newcommand{\pcl}{\mathbfit{L}}
\newcommand{\SO}{\mathbf{SO}}
\newcommand{\SPH}{\mathbf{S}}
\newcommand{\rotation}{\mathscr{R}}
\DeclareMathOperator*{\argmin}{\arg\!\min}

\journal{Annals of Nuclear Energy}

\makeatletter
\def\ps@pprintTitle{
 \def\@oddfoot{}%
 \let\@evenfoot\@oddfoot}
\makeatother

\begin{document}

\begin{frontmatter}

\title{Rotation-invariant Rapid TRISO-Fueled Pebble Identification Based on Feature Matching and Point Cloud Registration}

\author{Ming Fang}
\author{Angela {Di Fulvio}\corref{cor1}}
\cortext[cor1]{Corresponding author. Tel.: +1 217 300 3769.}
\ead{difulvio@illinois.edu}
\affiliation{organization={Department of Nuclear, Plasma, and Radiological Engineering, University of Illinois Urbana-Champaign},
            addressline={Suite 100 Talbot Laboratory, MC-234, 104 South Wright Street}, 
            city={Urbana},
            postcode={61801}, 
            state={IL},
            country={US}}

\begin{abstract}
Pebble bed reactor (PBR) relying on TRISO-fueled pebbles is one of the most promising Gen-IV reactor designs because of intrinsic safety and thermal efficiency. Fuel pebbles flow through PBR's core and the identification of individual pebbles exiting the core will be beneficial to improve safeguards and fuel management. We propose a pebble identification method that is fast, accurate, robust, and applicable to PBRs containing hundreds of thousands of pebbles. The identification relies on the internal distribution of TRISO fuel particles, which is a unique feature of each pebble. We experimentally demonstrated that X-ray CT can extract the particle distribution with high accuracy. We then applied the algorithm to identify a single pebble in a data set of 100,000 pebbles achieving 100\% identification accuracy in 90,000 tests with the presence of arbitrary rotations and measurement noises. The average time to identify one pebble is below \SI{50}{\s}, compatible with PBR operation.
\end{abstract}

\begin{keyword}
TRISO\sep PBR\sep rotation invariant\sep feature matching\sep point cloud registration\sep Go-ICP

\end{keyword}
\end{frontmatter}

\section{Introduction}

The next generation of nuclear reactors that feature enhanced safety, high thermal efficiency, and improved economics, are currently under active development worldwide~\cite{boll2023advanced,kelly2014generation,lorusso2018gen,pioro2022handbook}. Among different advanced reactor concepts, the pebble bed reactor (PBR) design is a promising candidate for wide deployment~\cite{kadak2005future,zhang2016shandong,mulder2021x}. Fuel elements in PBRs are typically \SI{6}{\cm} diameter graphite pebbles with approximately 10,000 coated Tri-Structural Isotropic (TRISO) fuel particles embedded. This form factor provides the benefits of high-temperature tolerance and effective retention of fission products~\cite{international2010high}. PBR's core is a vessel of hundreds of thousands of TRISO-fueled pebbles that are continuously circulating. When a pebble exits the core, burnup measurements are performed to determine if the pebble should be reinserted into the core for another cycle or be classified as spent and replaced by a fresh pebble. This online-fueling scheme has been shown to produce a more uniform burnup distribution compared to single-pass designs~\cite{topan2016study}.

Unlike LWR-type fuel, the fuel pebbles in PBRs have a much smaller form factor and are continuously flowing. This specific feature of PBRs calls for innovative fuel management, safeguards and material control \& accountability measures. A single fuel pebble contains a small amount of special nuclear material (SNM), i.e., $^{235}$U, and a large number of pebbles need to be diverted from the core for the SNM amount to reach a significant quantity (SQ)~\cite{forsberg2009safeguards}. {Under this consideration, item counting, in which number of pebbles exiting and entering the core are recorded and compared, and bulk analysis, in which a group of pebbles is collected into one container and inspected together, have been proposed to ensure no SQ of SNM is lost~\cite{kovacic2020advanced}. However, this approach alone is inadequate because it relies critically on the maintenance of continuity of knowledge (CoK) of fuel locations. Experiences on HTR-10 has shown that CoK can be lost and item counting fails to provide an accurate number of pebbles in the core~\cite{durst2009nuclear}. In this case, a measure must be in place to identify each individual fuel element to recover the CoK. In addition, broken pebbles in the core can not be tracked, which introduces materials unaccounted for (MUF). }Besides, it fails to address the potential consequence of losing a single spent fuel pebble that emits high-intensity radiation to the environment and anyone nearby.
It is therefore desirable to develop experimental techniques to enable identification (ID) of any individual pebble exiting the core. The ability to perform pebble identification also opens the door to exciting PBR research and reactor management at the scale of individual pebbles. The residence time and burnup change during a pebble's stay in the core can be inferred from its ID, which can be valuable for validation of computational fluid dynamics and neutronics codes. The number of passes and total residence time, inferred from the ID, can assist in the decision of pebble re-insertion or disposal. Knowledge of the pebble history and burnup throughout its lifespan will provide fuel designers and operators with valuable insight to inform safe and economical fuel handling protocols. 

Our identification method relies on the inherent and unique fingerprint associated with each fuel pebble. External identification marks on the pebbles' surfaces, similar to the serial number on LWR-type fuels, are not reliable as they can be easily reproduced and worn down because of pebble-pebble/pebble-wall frictions. Haire et al.~\cite{haire2007tags} suggested adding rare earth oxides to the uranium kernel as an internal tag, which requires modifications in fuel production and the identification can only be done destructively. Gitau et al.~\cite{gitau2012development} proposed randomly inserting ZrO${}_2$ particles of similar size as TRISO particles into the pebble and using the random placement of ZrO${}_2$ particles extracted through non-destructive ultrasound imaging as an identifier. However, this method also requires changes to the fuel fabrication process and the impact of ZrO${}_2$ spheres on the neutronics and structural strength needs to be investigated. Additionally, it would be technically challenging to identify a few ZrO${}_2$ particles among thousands of TRISO particles on the ultrasound image. Therefore, the inherent signature of the fuel pebbles, such as the 3D spatial distribution of fuel kernels, is determined to be more suitable for the identification purpose. The kernel distribution is unique for each fuel pebble due to the randomness introduced in pebble fabrication and is impossible to reproduce due to the large amount of kernels. X-ray Computed Tomography (CT) has been widely used for characterization of TRISO-fueled compacts for quality control, and the TRISO particles can be easily resolved due to their higher density compared to the surrounding graphite~\cite{helmreich2020new,vrinda2019triso,kane20223d,yu20173d}. Kwapis et al.~\cite{KWAPIS2021103913} developed a neural-network algorithm to perform pebble identification based on the X-ray CT projections of the pebble. However, the algorithm was not able to identify spherical pebbles with rotations beyond $2.5^\circ$. This maximum displacement angle limits the applicability of this method.
In this work, we build upon our previous work that demonstrated the feasibility of identifying a single pebble in the presence of arbitrary rotations and high measurement noise, in a relatively small dataset of 100 pebbles ~\cite{fang2022algorithms}. In this work, we propose the first TRISO-fueled pebble identification algorithm to achieve accurate and robust identification of a single pebble in a library of 100,000 pebbles in{ less than \SI{50}{\s}}. 

This paper is organized as follows. In Section~\ref{sec:experimental_method}, we demonstrate the experimental extraction of the spatial distribution of high-density kernels through X-ray CT for identification purposes. In Section~\ref{sec:computational_method}, we introduce the pebble identification algorithm based on rotation-invariant feature matching and point cloud registration. In Section~\ref{sec:results}, we apply the identification algorithm to a dataset of 100,000 pebbles and achieve 100\% accuracy in 10,000 tests. Finally, the discussion and conclusions are presented in Section~\ref{sec:conclusion}.

\section{Experimental Methods}\label{sec:experimental_method}
In this section, we experimentally demonstrated the extraction of the spatial distribution of high-density kernels from a mock-up fuel compact with an industrial X-ray CT scanner.{ We compared the reconstructed kernel distribution to the ground truth and calculated the fraction of outliers, which includes fractions of both false positives (kernels in our reconstruction that do not have a correspondence in the ground truth), and vice versa, for the false negatives. Based on the outlier rates, we generated synthetic test datasets by adding random outliers in Section~\ref{sec:computational_method} for testing of pebble identification algorithms.}

\subsection{Cone-Beam X-ray CT Scan of Mock-up Fuel Compact}
\begin{figure}[!htbp]
	\centering
	\includegraphics[width=.8\linewidth]{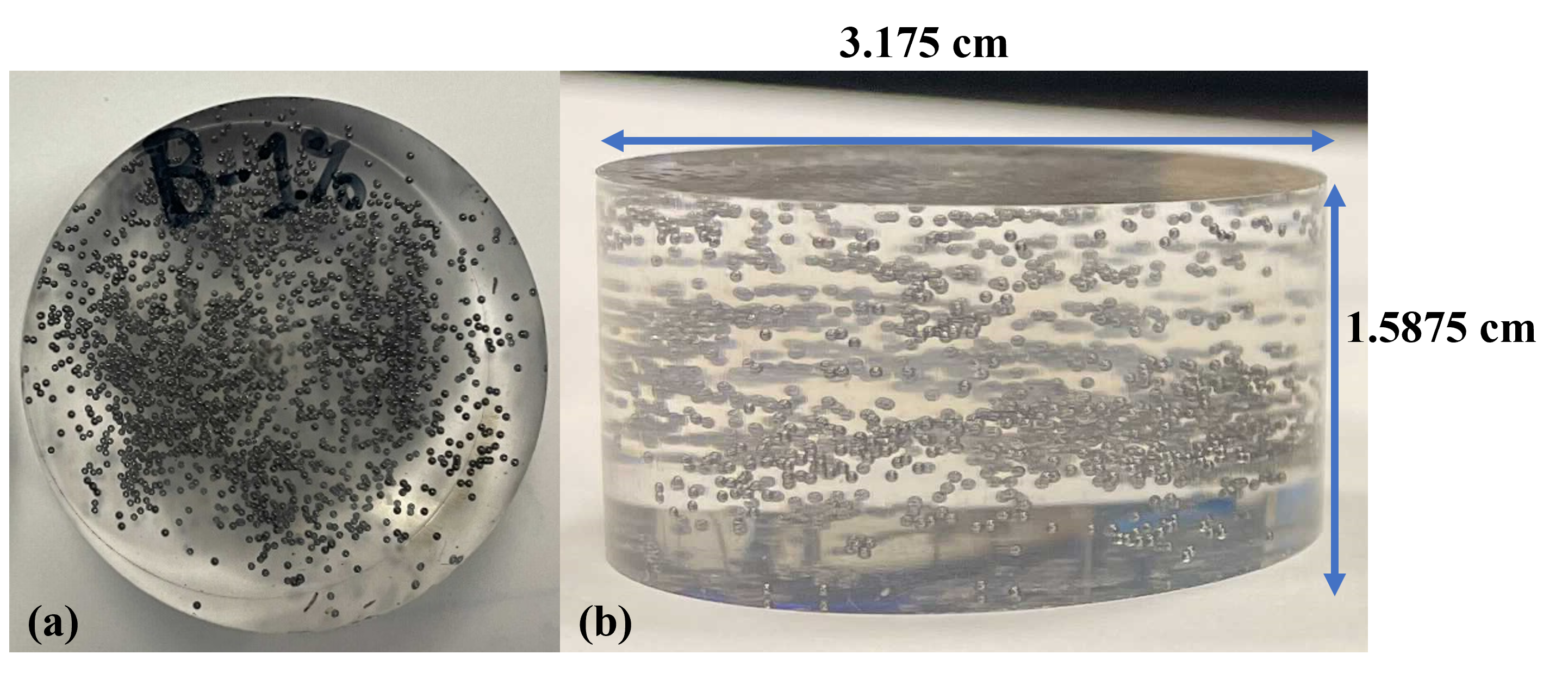}
	\caption {(a) Top and (b) side picture of a mock-up fuel compact sample provided by BWX Technologies, Inc. (BWXT) made of approximately 2000 \SI{500}{\micro\meter} diameter tungsten-carbide (WC) kernels (black) and lucite plastic (transparent).}
	\label{fig:mockup_fuel}
\end{figure}

A mock-up fuel sample was made by mixing tungsten-carbide (WC) kernels of \SI{500}{\um} diameter with Lucite thermoplastic metallographic mounting material (LECO 811-132, chemical formula (C\textsubscript{5}H\textsubscript{8}O\textsubscript{2})\textsubscript{n}). Fig.~\ref{fig:mockup_fuel} shows the WC sample with 1\% WC volume loading fraction. The material composition is shown in Table~\ref{table:mat_composition}. 

\begin{table}[!htbp]
 \footnotesize
\centering
\caption{Material composition of the WC-loaded sample and a 3-cm radius TRISO-fueled pebble with 10,000 fuel particles~\cite{terry2005evaluation}.}\label{table:mat_composition}
\begin{tabular}{|c|ccc|ccc|}
\hline
\multirow{2}{*}{}         & \multicolumn{3}{c|}{TRISO-fueled Pebble}                                 & \multicolumn{3}{c|}{WC-loaded sample}                                    \\ \cline{2-7} 
                          & \multicolumn{1}{c|}{Material} & \multicolumn{1}{c|}{Thickness} & Density & \multicolumn{1}{c|}{Material} & \multicolumn{1}{c|}{Thickness} & Density \\ \hline
Kernel                    & \multicolumn{1}{c|}{UO\textsubscript{2}}      & \multicolumn{1}{c|}{\SI{500}{\um}}       & \SI[per-mode=symbol]{10.4}{\gram\per\cubic\cm}    & \multicolumn{1}{c|}{WC}       & \multicolumn{1}{c|}{\SI{500}{\um}}       & \SI[per-mode=symbol]{15.63}{\gram\per\cubic\cm}   \\ \hline
\multirow{4}{*}{Coatings} & \multicolumn{1}{c|}{Buffer} & \multicolumn{1}{c|}{\SI{90}{\um}}        & \SI[per-mode=symbol]{1.1}{\gram\per\cubic\cm}     & \multicolumn{3}{c|}{\multirow{4}{*}{None}}                               \\ \cline{2-4}
                          & \multicolumn{1}{c|}{IPyC} & \multicolumn{1}{c|}{\SI{40}{\um}}        & \SI[per-mode=symbol]{1.9}{\gram\per\cubic\cm}     & \multicolumn{3}{c|}{}                                                    \\ \cline{2-4}
                          & \multicolumn{1}{c|}{SiC}      & \multicolumn{1}{c|}{\SI{35}{\um}}        & \SI[per-mode=symbol]{3.18}{\gram\per\cubic\cm}    & \multicolumn{3}{c|}{}                                                    \\ \cline{2-4}
                          & \multicolumn{1}{c|}{OPyC} & \multicolumn{1}{c|}{\SI{40}{\um}}        & \SI[per-mode=symbol]{1.9}{\gram\per\cubic\cm}     & \multicolumn{3}{c|}{}                                                    \\ \hline
Matrix                    & \multicolumn{1}{c|}{Graphite} & \multicolumn{1}{c|}{-}         & \SI[per-mode=symbol]{1.73}{\gram\per\cubic\cm}    & \multicolumn{1}{c|}{Lucite}   & \multicolumn{1}{c|}{-}         & \SI[per-mode=symbol]{1.18}{\gram\per\cubic\cm}    \\ \hline
\# of kernels       & \multicolumn{3}{c|}{10000}                                               & \multicolumn{3}{c|}{$\sim$2000}                                         \\ \hline
\end{tabular}
\end{table}

\begin{figure}[!htbp]
	\centering
	\includegraphics[width=.5\linewidth]{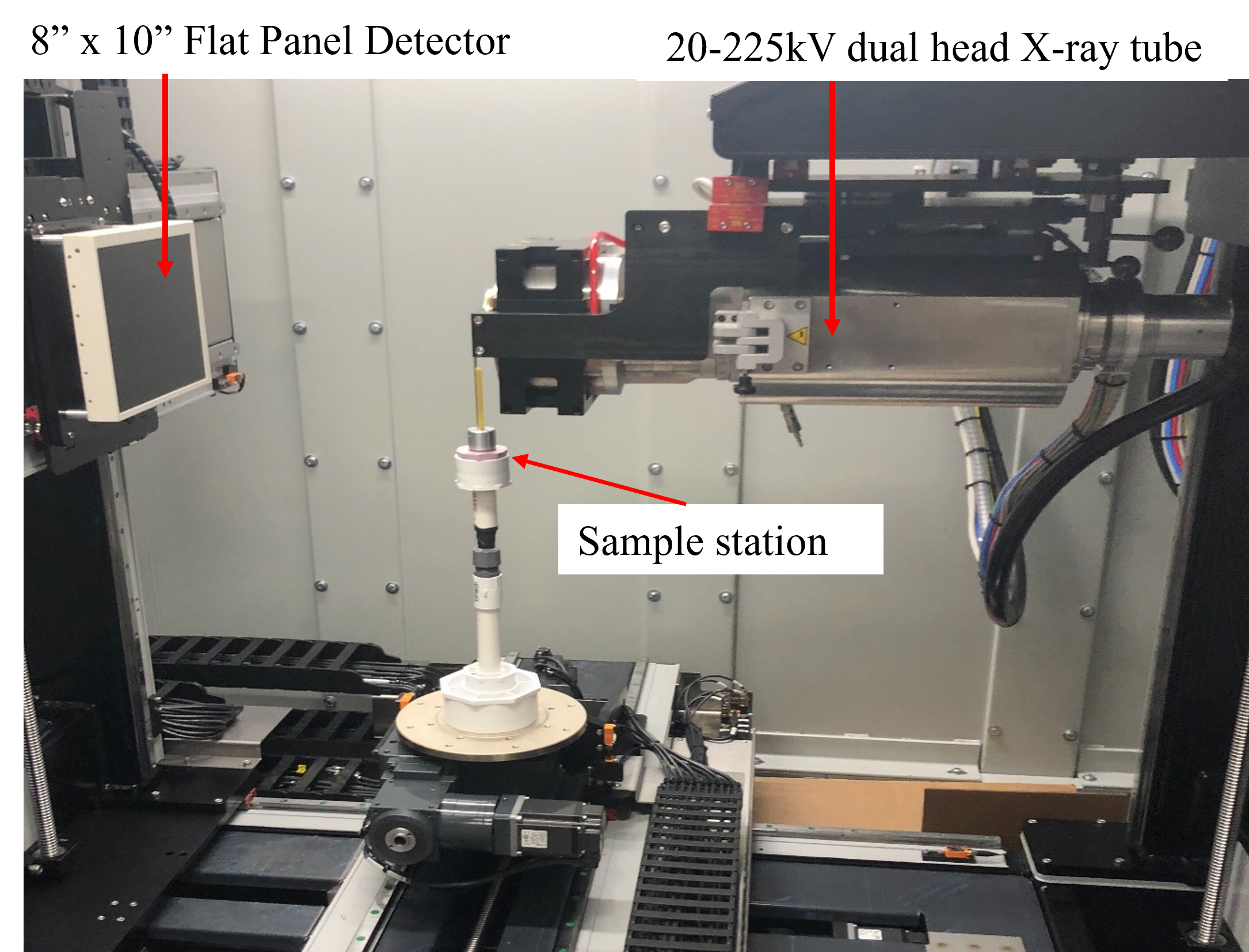}
	\caption{Scanning area of the NSI X5000 CT system.}
	\label{fig:xray-setup}
\end{figure}

We performed cone-beam X-ray CT scan of the WC-loading sample using a North Star Imaging X5000 industrial CT scanner. Figure~\ref{fig:xray-setup} shows the scanning system. During the CT scan acquisition, the X-ray tube voltage was set to 200 kVp and the current was \SI{36}{\uA}. The flat panel detector is Varex PaxScan 2520DX consisting of $1536\times1920$ pixels, with a pixel pitch of \SI{127}{\um}. The frame rate is 12.5~fps at full resolution and 30~fps at 768$\times$960 resolution. The source-to-detector distance was \SI{301.277}{\mm} and the source-to-object distance was \SI{54.875}{\mm}, resulting in a zoom factor of 5.49 and an effective voxel pitch of \SI{23.13}{\um}. The angle increment was ${1}^{\circ}/{7}$ and a total of 2520 projections were acquired and saved as 16-bit tiff images.

\begin{figure}[!htbp]
    \begin{subfigure}[t]{0.5\linewidth}
        \centering
        \includegraphics[width=.95\linewidth]{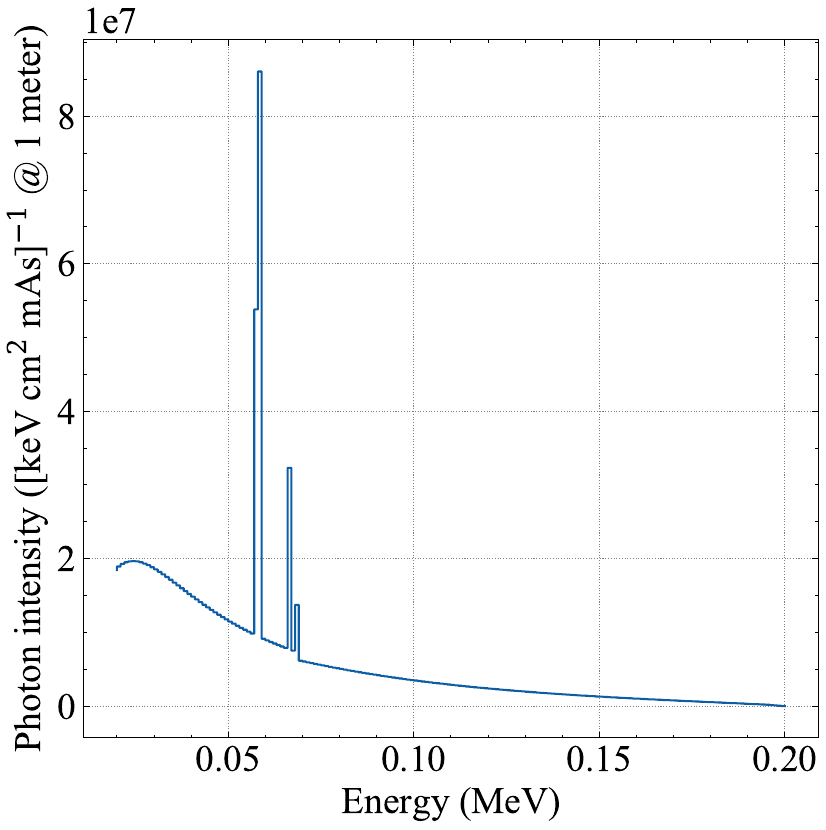}
        \caption{200kVp X-ray spectrum}\label{fig:200kVp_spectrum}
    \end{subfigure}\hfil
    \begin{subfigure}[t]{0.5\linewidth}
        \centering
        \includegraphics[width=.95\linewidth]{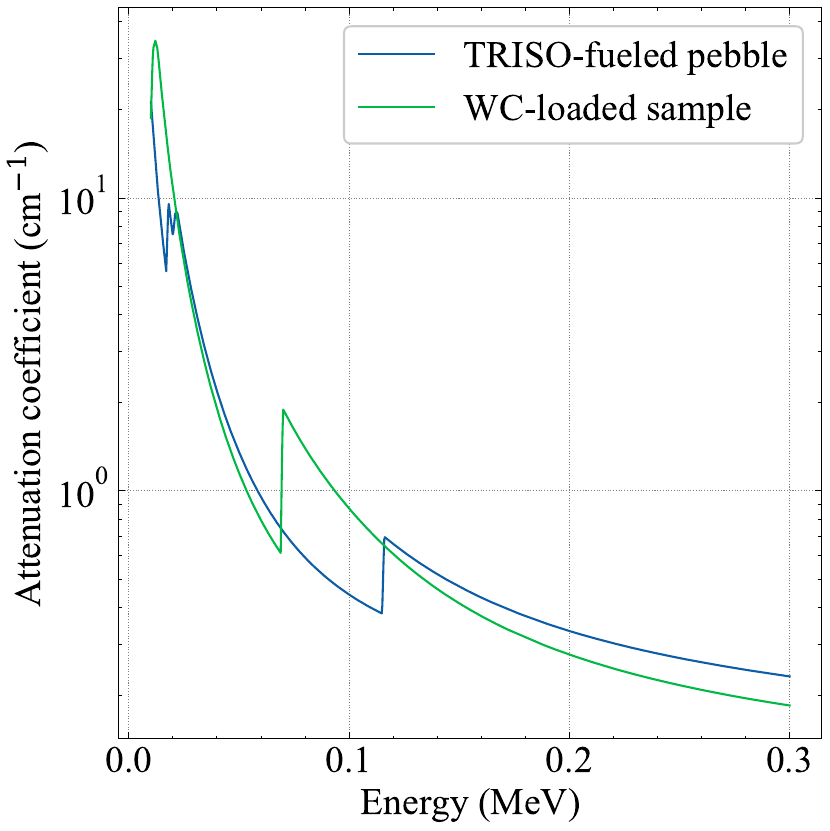}
        \caption{Attenuation coefficient}\label{fig:comp_atten_coeff}
    \end{subfigure}
	\caption{(a) Energy spectrum of 200~kVp X-rays calculated using SpekCalc~\cite{Poludniowski2009}. (b) Energy-dependent attenuation coefficient of the homogenized WC-loaded sample and TRISO-fueled pebble.}
\end{figure}

{The X-ray imaging method described here can be applied to TRISO-fueled pebbles, provided that X-ray transmission coefficients of the WC-loaded sample and a TRISO-fueled pebble are comparable. }Figure~\ref{fig:200kVp_spectrum} shows the energy spectrum of 200~kVp X-rays, with characteristic X-ray peaks in the \SI{50}{\kilo\eV}-\SI{100}{\kilo\eV} energy range superimposed to the bremsstrahlung continuum.{ The characteristic peaks result from the tungsten anode of the X-ray tube.} We first calculated the attenuation coefficients of the homogenized WC-loaded sample and TRISO-fueled pebble based on Table~\ref{table:mat_composition}. As shown in Fig.~\ref{fig:comp_atten_coeff}, the WC-loaded sample exhibits a higher attenuation coefficient in the \SI{50}{\kilo\eV}-\SI{100}{\kilo\eV} energy range, due to the K-edge of tungsten, which compensates for the smaller radius. We used MCNP (Monte Carlo N-Particle)~\cite{osti_1419730} to simulate a 200~kVp parallel X-ray beam impinging on the WC-loaded sample and the TRISO-fueled pebble. The average X-ray transmission coefficients of the WC-loaded sample and the TRISO-fueled pebble are 11.7\% and 19.5\%, respectively, which are comparable.

\subsection{Image Reconstruction}
The image reconstruction was performed using the All Scale Tomographic Reconstruction Antwerp (ASTRA) toolbox~\cite{van2016fast}, an image reconstruction toolbox with GPU acceleration. The reconstruction process consists of two steps. In the first step, we pre-processed the data to remove any artifacts on the 2-D projections. We subtracted the background counts from the projections, and corrected for the variation in pixel gains:
\begin{equation}
    I' = \frac{I-D}{F-D}
\end{equation}
where $I$ and $I'$ are the image projections before and after the correction, $D$ is the dark field and $F$ is the flat field. We then linearized the data:
\begin{equation}
    I'' = -\log(I')
\end{equation}

In the second step, we employed the GPU-accelerated FDK (Feldkamp-Davis-Kress) algorithm~\cite{feldkamp1984practical} in the ASTRA toolbox to reconstruct a 3-D image from the linearized projection data $I''$. The cone-beam scan geometry was provided as an input to the reconstruction algorithm.{ To speed up reconstruction, the X-ray projection was down-sampled into $768\times 960$ pixels and 360 projections were used in the reconstruction. The resulting voxel pitch was \SI{46}{\um}.}

For comparison, we also performed image reconstruction using the software provided by NSI. { The NSI reconstruction utilized all the 2520 scans with no down-sampling and the voxel pitch was \SI{23}{\um}. The NSI reconstruction was used as the ground truth, which we compared the ASTRA-reconstruction to for evaluating the reconstruction quality.}

\subsection{Image Segmentation}
Given a reconstructed image, our next step is to identify the WC kernels inside the sample and determine their locations through image segmentation. First, we applied Ostu's method~\cite{otsu1979threshold} to classify the pixels into two categories based on their values, considering one class as the WC kernels and the other as the surrounding plastic. We then applied a white top-hat filter to remove any reconstruction artifacts on the binary image, followed by a Laplacian-of-Gaussian-based blob detection algorithm to identify the cross sections of the WC kernels on the binary image. Implementations in the \texttt{scikit-image} package were used in this step~\cite{scikit-image}. We repeated these steps for all slices of the 3D reconstruction and finally, the kernels' 3D positions were found by merging the 2D blobs.

We applied image segmentation to the NSI-reconstructed image and used the extracted kernel distribution as the ground truth. We compared the kernel distribution based on ASTRA-reconstruction to the ground truth and calculated the false positive (FP) rate, which is defined as the fraction of kernels in our segmentation that do not have a correspondence in the ground truth, and vice versa, for the false negative rate (FN) rate. A correspondence between a pair of points is established if the distance between them is less than \SI{500}{\um} - the diameter of a single WC kernel.

\section{Computational Methods}\label{sec:computational_method}

\subsection{Overview of Pebble Identification Algorithm}
According to our pebble identification method, in a PBR with an online-refueling scheme, a pebble exiting the reactor core will be scanned by a X-ray CT scanner and the spatial coordinates of the uranium-bearing kernels will be extracted through image processing. The set of all kernels' 3D spatial coordinates is called a point cloud. We assume a pebble is uniquely determined by the corresponding point cloud (up to rotations), which is the basis of our identification concept, and we will use the two terms interchangeably in this section. The other input to the algorithm is the set of point clouds of all existing fuel pebbles that have been scanned previously, referred to as the library. The number of pebbles in the library depends on the design of the PBR and is usually on the order of 100,000. We are tasked to determine whether the pebble under inspection is one of the existing pebbles in the library and extract the corresponding ID if so.

A successful pebble identification algorithm should meet the following requirements:
\begin{enumerate}
    \item Fast: Retrieve the pebble ID from a library of 100,000 pebbles within a few minutes to ensure a continuous flow of pebbles through the core.
    \item Rotation-invariant and permutation-invariant: Retrieve the correct pebble ID regardless of the orientation of the pebble and the order of the points in the point cloud.
    \item Robust against noises of kernels' positions and outliers (missing/deformed kernels and non-existing kernels) introduced during image processing.
\end{enumerate}

Let $\pcx=\{\vecx_j\in \mathbb{R}^3|1\leq j \leq n\}$ stand for the point cloud of the pebble under inspection, where  $\vecx_j$ is the spatial coordinate of kernel $j$ and $n$ the total number of kernels. $\pcl_K=\{\pcy_1, \pcy_2, \cdots, \pcy_K\}$ is the library of pebbles, where $\pcy_i=\{\vecy_j\in \mathbb{R}^3|1\leq j \leq n_i\}$ is the point cloud of pebble with ID$=i$ and $K$ the total number of pebbles in the library, which is set to 100,000 in this study.

In this work, we propose a coarse-to-fine approach to enable fast and accurate pebble identification in PBRs, outlined in Algorithm~\ref{alg_1:identification_outline}. In Step 1 (coarse search), we compare rotation-invariant features of the input point cloud and library, which allows us to reduce the search space from $K=100,000$ pebbles to $N=100$ pebbles. After executing Step 1, a reduced-size library $\pcl_N$ of $N$ pebbles is generated, which is fed to Step 2. In Step 2 (fine search), we calculate the difference between the input point cloud and any point cloud in $L_N$ and retrieve the ID that gives the smallest difference. The key of Algorithm~\ref{alg_1:identification_outline} are the two metric functions $d_1(\pcx,\pcy)$ and $d_2(\pcx,\pcy)$, which measure the difference between two point clouds $\pcx$ and $\pcy$. To meet the above-mentioned requirements, $d_1(\pcx,\pcy)$ should be relatively fast to evaluate, and both $d_1(\pcx,\pcy)$ and $d_2(\pcx,\pcy)$ should be rotation-invariant, permutation-invariant and robust against noises. We will discuss them in detail in Section~\ref{ssec:search_space_reduction} and Section~\ref{ssec:GoICP-based_id_retrieval}.

\begin{figure}[!htbp]
	\centering
	\includegraphics[width=\linewidth]{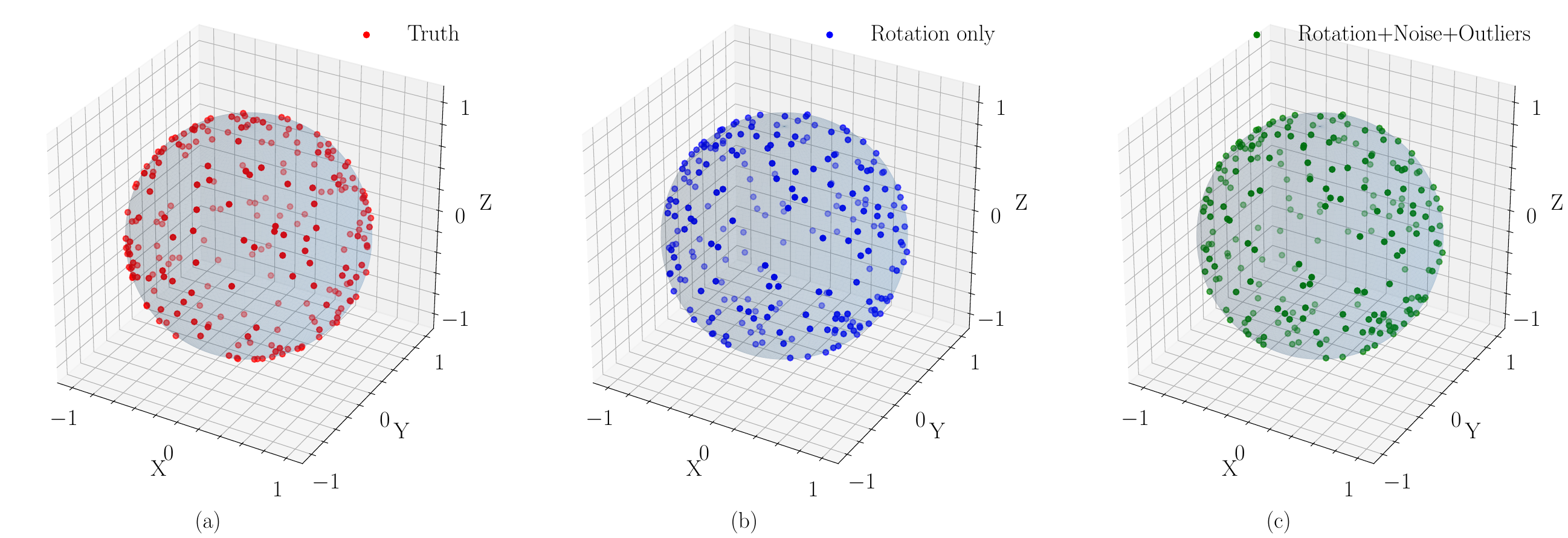}
	\caption{(a) A point cloud containing 244 points, (b) the randomly rotated point cloud, (c) the randomly rotated point cloud with Gaussian noises and outlier points added. The points are normalized to have norms of 1.}
	\label{fig:library_test_example}
\end{figure}

A simplification we have made in constructing the point cloud is to select only the hundreds of points in a spherical shell close to the non-fuel zone, i.e., the outermost layer. This choice is due to three reasons. First, we found that these superficial points carry sufficient information about the pebble to enable unique identification, as shown in the Results. Second, the identification time grows quadratically with the number of points, and using all the points is time-consuming. Last also most importantly, these points are the easiest to image and locate by conventional X-ray CT scanners with moderate penetration depth{ due to the lower average energy}, and the measurement noises are expected to be lower near the surface. These superficial points are then projected onto the surface of the unit sphere to form a normalized point cloud, which is used for identification. Figure~\ref{fig:library_test_example} shows the comparison of a normalized point cloud and its rotated version with position noises and outlier points added.

\begin{algorithm}[H]
    \caption{Two-step Pebble Identification}\label{alg_1:identification_outline}
    \begin{algorithmic}[1]
        \Procedure{Step 1: Search space reduction}{}\newline
        \textbf{Input:} Pebble $\pcx$. Library of $K$ pebbles $\pcl_K=\{\pcy_i|i\leq K\}$. Reduced library size $N$.
        \For{each $\pcy_i$}
            \State{Calculate the rotation-invariant features of $\pcx$ and $\pcy_i$}
            \State{Calculate the difference between the features: $d_i = d_1(\pcx,\pcy_i)$}
            \State{Append $d_i$ to the list of differences}
       \EndFor
       \State{Sort the list of differences.}
        \State{\textbf{Return:} $N$ pebbles with the smallest differences.}
        \EndProcedure
        \Procedure{Step 2: Pebble ID Retrieval}{}\newline
        \textbf{Input:} Pebble $\pcx$. Library of $N$ pebbles $\pcl_N=\{\pcy_i|i\leq N\}$. Threshold $\epsilon$.
        \For{each $\pcy_i$}
            \State{Calculate the difference between $\pcx$ and $\pcy_i$: $d_i = d_2(\pcx,\pcy_i)$}
            \State{Append $d_i$ to the list of differences}
       \EndFor
       \State{Calculate the minimum value of the list of differences, $d_{\min}$}
        \If{$d_{\min}<\epsilon$}
          \State{\textbf{Return:} The ID corresponding to the minimum difference}
        \Else
          \State{\textbf{Return:} No ID is found. Append this new pebble to $L_K$}
      \EndIf 
        \EndProcedure
    \end{algorithmic}
\end{algorithm}

\subsection{Search Space Reduction Based on Rotation-invariant Feature Matching}\label{ssec:search_space_reduction}
Comparing the input pebble $\pcx$ with each pebble $\pcy$ in the library is time-consuming due to the large number of candidates. In this section, we seek to find a metric function $d_1(\pcx,\pcy)$ that measures the difference between two point clouds and rejects $\pcy$ if $d_1(\pcx,\pcy)$ is large. We will show that this approach can reduce the search space from 100,000 pebbles to 100 pebbles. Our main goal is to construct a descriptor function $\mathcal{H}$ that extracts a rotation-invariant feature $\mathcal{H}(\pcx)$ from the given point cloud $\pcx$. If the features $\mathcal{H}(\pcx)$ and $\mathcal{H}(\pcy)$ are very different, we conclude that $\pcx\neq\pcy$.

\begin{figure}[!htbp]
	\centering
	\includegraphics[width=.5\linewidth]{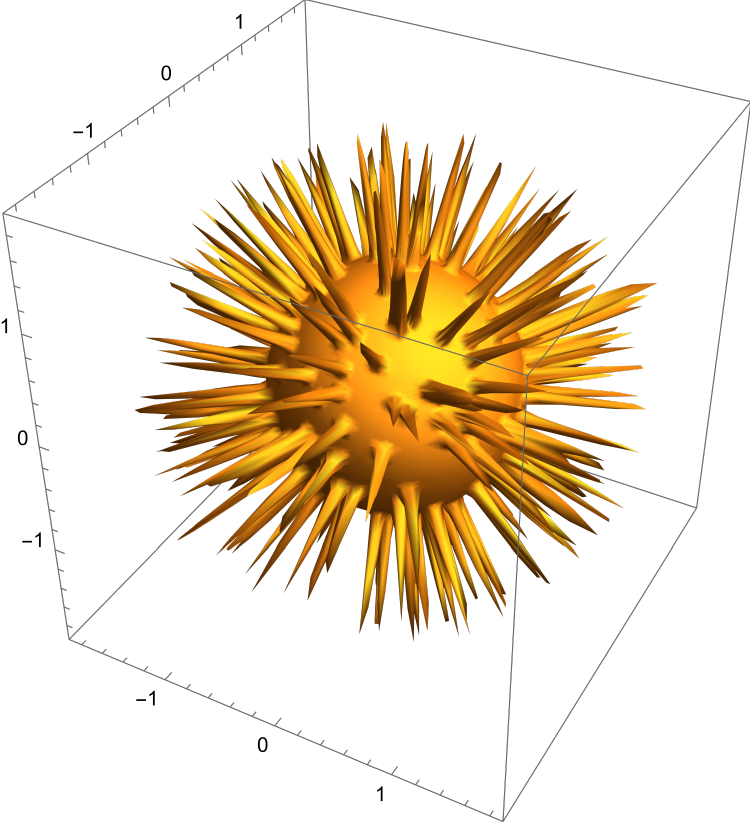}
	\caption{A 3D surface plot of the spherical function $f(\theta,\phi)$ corresponding to the point cloud in Fig.~\ref{fig:library_test_example}(a) when $\epsilon=0.02$. The distance to the unit sphere represents the function value.}
	\label{fig:spherical_func_3d}
\end{figure}

The rotation-invariant feature extraction process consists of two steps. First, given any point cloud $\pcx=\{\vecx_i|\vecx_i\in \SPH^2,1\leq i\leq n\}$, we define a continuous and square-integrable function $f$ on the unit sphere $\SPH^2$:
\begin{gather}
    f(\theta, \phi)=\sum_{i=1}^n f_i(\theta, \phi), f_i(\theta, \phi)=\frac{1}{\sqrt{\pi}\epsilon}\exp(-\frac{\|\vecx-\vecx_i\|^2}{2\epsilon^2}),\label{eq:sph_fuc_def}\\
    \vecx=(\sin\theta\cos\phi,\sin\theta\sin\phi,\cos\theta)\in\SPH^2, 0\leq\theta\leq\pi,-\pi\leq\phi\leq\pi
\end{gather}
$f_i(\theta,\phi)$ is a Gaussian-like function and attains its maximum at point $\vecx_i$. Permutation-invariance is guaranteed by the summation of all points. $\epsilon$ controls the spread of the Gaussian kernel, and as $\epsilon$ approaches zero, $f_i$ becomes delta-like, as shown in Fig.~\ref{fig:spherical_func_3d}. Figure~\ref{fig:spherical_func_3d} also shows that it is possible to reconstruct $\pcx$ from $f(\theta,\phi)$ by enumerating all local maxima of $f(\theta,\phi)$, therefore no information loss is introduced by our method. In this study, we assumed a fuel particle diameter of \SI{910}{\um}~\cite{terry2005evaluation} and the minimum angular distance between two points in the pebble is \SI{910}{\um}/\SI{25}{mm}=\SI{36.4}{\milli\radian}. Therefore, we set $\epsilon$ to \SI{40}{\milli\radian} to ensure good separation between the points.

The second step is to extract the power spectrum of $f(\theta,\phi)$, which is our desired rotation-invariant feature~\cite{kazhdan2003rotation}. It is known that a spherical function $f(\theta,\phi)$ can be decomposed into spherical harmonics:
\begin{equation}
    f(\theta,\phi) = \sum_{l=0}^{+\infty}\sum_{m=-l}^l a_{lm}Y_{lm}(\theta,\phi) 
\end{equation}
where $Y_{lm}(\theta,\phi)$ is the spherical harmonic function of degree $l$ and order $m$. The expansion coefficient $a_{lm}$ is given by: 
\begin{equation}
    a_{lm}=\langle Y_{lm},f(\theta,\phi)\rangle = \int_{\phi=-\pi}^{\pi}\int_{\theta=0}^{\pi} f(\theta,\phi)Y_{lm}^*(\theta,\phi)\sin\theta d\theta d\phi
\end{equation}
The rotation-invariant feature $h\coloneqq \mathcal{H}(\pcx)$ is defined as the $L_2-$norm spectrum of $f(\theta,\phi)$:
\begin{equation}
    h=\{\|f_l\| | l\geq 0\}=\left\{\sqrt{\sum_{m=-l}^l |a_{lm}|^2} \Bigg| l\geq 0\right\}, f_l=\sum_{m=-l}^{l}a_{lm}Y_{lm}(\theta,\phi)\label{eq:feature_definition}
\end{equation}
The rotation-invariance property is proved in \ref{apdx:rotation_invariance}. 

Finally, based on the descriptor $\mathcal{H}$, we define the metric function $d_1(\pcx, \pcy)$ in Algorithm~\ref{alg_1:identification_outline} as the sum of squared-differences between their features:
\begin{equation}
    d_1(\pcx, \pcy)\coloneqq\|\mathcal{H}(\pcx)-\mathcal{H}(\pcy)\|^2=\sum_{l=0}^{\infty}\|f_l(\pcx)-f_l(\pcy)\|^2
\end{equation}
Given an input pebble $\pcx$, we reject a pebble $\pcy_j$ from the library that results in large $d_1(\pcx, \pcy_j)$, as they cannot be the same pebble, therefore achieving an important reduction of search space. The features of existing pebbles $\{\mathcal{H}(\pcy_j)|j\leq K\}$ can be pre-computed and stored in the library and updated when a new pebble is inserted or a spent pebble is removed.

\begin{figure}[!htbp]
	\centering
	\includegraphics[width=\linewidth]{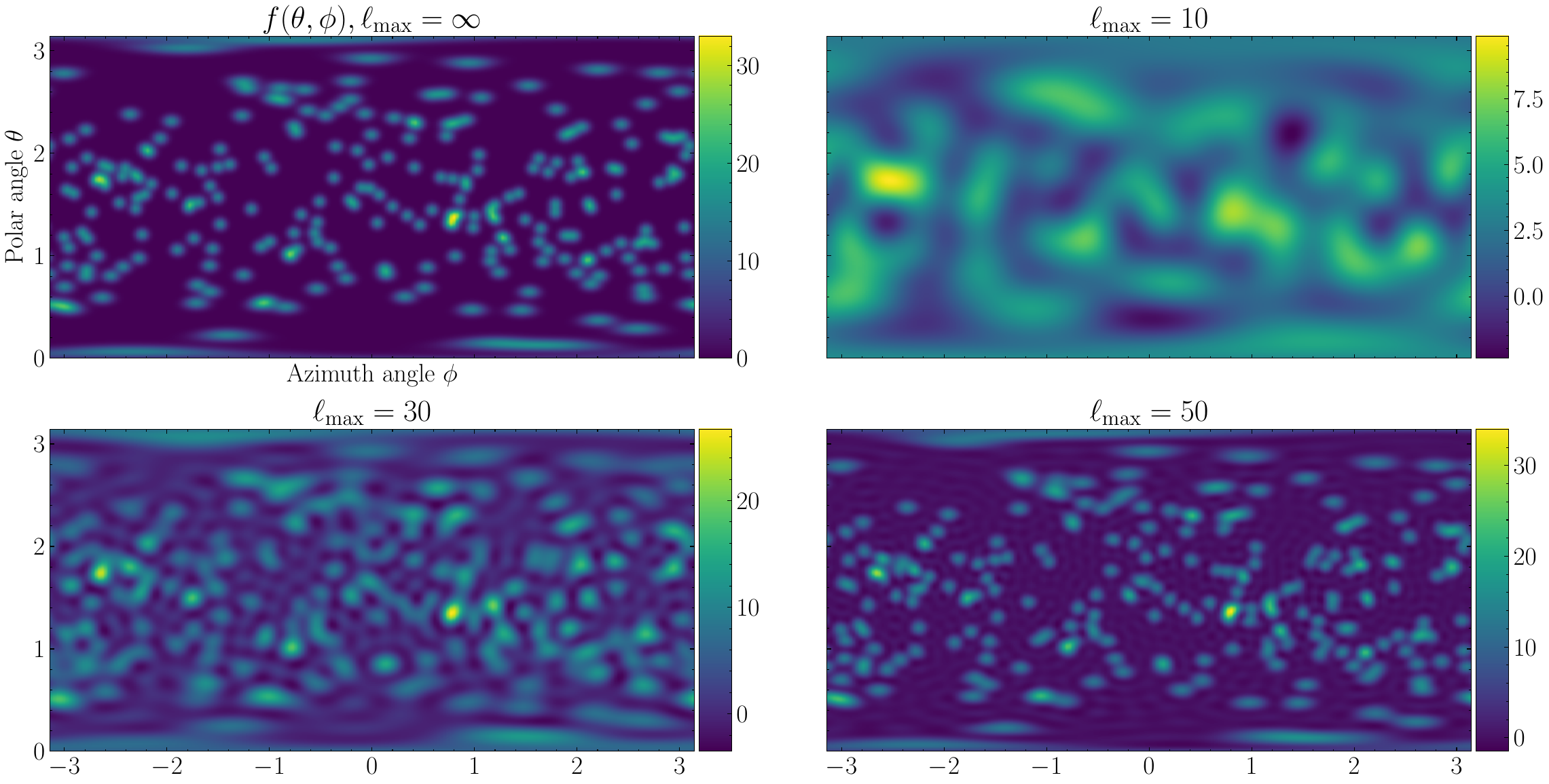}
	\caption{2D plot of the spherical function $f(\theta, \phi)$ corresponding the point cloud in Fig.~\ref{fig:library_test_example}(a), and its spherical harmonics expansion truncated at maximum degrees of 10, 30, and 50.}
	\label{fig:sphe_harm_expan_lmax}
\end{figure}

\begin{figure}[!htbp]
	\centering
	\includegraphics[width=.5\linewidth]{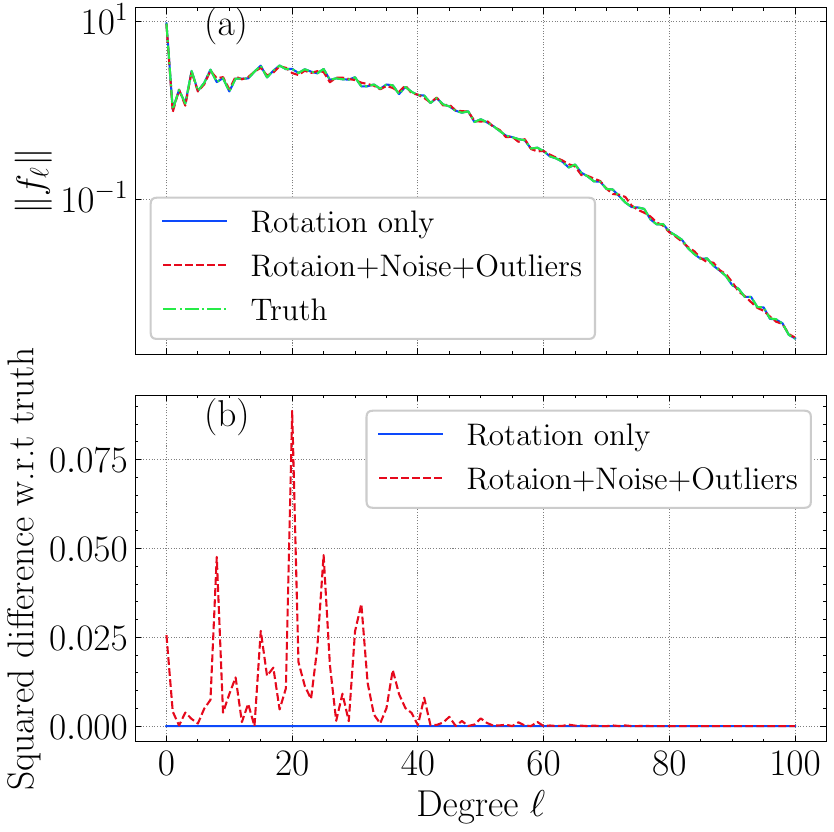}
	\caption{Comparison of rotation-invariant features of the three point clouds in Fig.~\ref{fig:library_test_example}.}
	\label{fig:comp_descriptor}
\end{figure}

In numerical implementation, we used the \texttt{SHTOOLS} library to achieve fast evaluation of the rotation-invariant feature~\cite{mark_wieczorek_2019_3457861,wieczorek2018shtools}. It's also necessary to truncate the $L_2$-norm spectrum at a maximum degree $\ell_{\max}$ without losing too much information. Figure~\ref{fig:sphe_harm_expan_lmax} shows the spherical function of $f(\theta,\phi)$ corresponding to the point cloud in Fig.~\ref{fig:library_test_example}(a), as well as the spherical harmonics expansion truncated at $\ell_{\max}=10,30,50$. We observe that for a point cloud containing 200-300 points, a maximum degree of $50$ is sufficient, which is used in this study.{ Figure~\ref{fig:comp_descriptor} compares the features extracted from the three point clouds in Fig.~\ref{fig:library_test_example}.} One may notice that the extracted feature does not change with rotations and the additional noises and outliers result in small differences. The first term at $l=0$ is not used when calculating $d_1(\pcx,\pcy)$ because although it is the dominant term of the spectrum, as shown in Fig.~\ref{fig:comp_descriptor}, it mainly depends on the number of points and including $f_0$ will make the algorithm less sensitive to the change of point distribution.

It should be noted that the summation over $m$ in Eq.~\eqref{eq:feature_definition} leads to information loss and it is possible to have two different pebbles with the same feature, i.e., $\mathcal{H}(\pcx)=\mathcal{H}(\pcy)$ does not imply $\pcx=\pcy$. For example, we can prove that the descriptor $\mathcal{H}$ is reflection-invariant (\ref{apdx:reflection_invariance}), i.e., a pebble and its reflection about any 2D plane, being not the same pebble, will result in the exact same feature. This ambiguity due to information loss will be addressed in the next section.

\subsection{Identification Based on Point Cloud Registration}~\label{ssec:GoICP-based_id_retrieval}
In the previous section, we introduced a coarse search to reduce the size of the search space. The coarse search algorithm retrieves a list of pebbles for which $\pcl_N=\{\pcy_i|i \leq N\}$. We now seek to perform a fine search on this list and find the one matching the input pebble $\pcx$. It should be noted that the matching pebble may not be the first element of the list of the coarse algorithm due to noises and outliers. What's more, because of the information loss during feature extraction, two different fuel pebbles can have the same rotation-invariant features, in which case the coarse-search algorithm is unable to discriminate between them. For these reasons, it is necessary to loop through $\pcl_N$ and compare each element to $\pcx$. 

In this section, we introduce a pebble identification algorithm based on point cloud registration. Point cloud registration is an important research problem in 3D computer vision, in which one tries to align a reference point cloud $\pcy$ with the input point cloud $\pcx$ through spatial transformations (e.g., scaling, rotation and translation)~\cite{huang2021comprehensive}. In this problem, the transformation is restricted to rotation only. For each pebble $\pcy$ in the library, we align $\pcy$ with $\pcx$ using point cloud registration, and retrieve the pebble ID that yields the minimum difference after alignment. 

Formally, given a reference point cloud $\pcy=\{\vecy_j|1\leq j\leq m\}$ and an input point cloud $\pcx=\{\vecx_k|1\leq k\leq n\}$, we define a cost function as a function of rotation $\rotation$:
\begin{gather}
    J(\rotation,\pcx,\pcy)=\frac{1}{m}\sum_{j=1}^m d(\rotation\vecy_j, \pcx)^2,\rotation\in\SO(3)\label{eq:cost_func}\\
    d(\rotation\vecy_j, \pcx)=\min_k \|\vecx_k-\rotation\vecy_j\|\label{eq:opt_1}
\end{gather}
where $\SO(3)$ is is the group of all rotations about the origin of 3D Euclidean space $\mathbb{R}^{3}$, $d(\rotation\vecy_j, \pcx)$ is the minimum distance between a point of index $j$ in $\rotation\pcy$ and all points in $\pcx$, $J$ is the mean-squared-error (MSE). We calculate the global minimum of the cost function and define it as the metric $d_2$ in Algorithm~\ref{alg_1:identification_outline}:
\begin{equation}
    d_2(\pcx,\pcy)\coloneqq \underset{\rotation\in \SO(3)}{\min}~J(\rotation, \pcx, \pcy)\label{eq:opt_2}
\end{equation}
The permutation-invariance and rotation-invariance of $d_2(\pcx,\pcy)$ are guaranteed by the summation in Eq.~\eqref{eq:cost_func} and the minimization in Eq.~\eqref{eq:opt_2}, respectively. 
Finally, we find the pebble ID by minimizing $d_2(\pcx,\pcy_i)$ over all possible $i$:
\begin{equation}
    \mathrm{ID} = \underset{i\geq 1}{\argmin}~d_2(\pcx,\pcy_i)\label{eq:opt_3}
\end{equation}
and reject the ID if $d_2(\pcx,\pcy_{\mathrm{ID}})$ is greater than a preset-threshold $\epsilon$, in which case no matching pebble ID is found. 

As shown in Eqs.~\eqref{eq:opt_1},\eqref{eq:opt_2},\eqref{eq:opt_3}, there are three nested layers of optimization involved in determining the ID of a single pebble. The algorithm is therefore time-consuming and becomes impractical when the size of the library is large, which necessitates the size reduction in the previous section. In the following, we briefly describe the strategies for solving the optimization problems in Eqs.~\eqref{eq:opt_1} and \eqref{eq:opt_2}. The outermost optimization Eq.~\eqref{eq:opt_3} is solved by simply looping through all $i$. 

\subsubsection{Solving Eq.~\eqref{eq:opt_1}{ to find the nearest neighbour distance}}
\begin{figure}[!htbp]
	\centering
	\includegraphics[width=.5\linewidth]{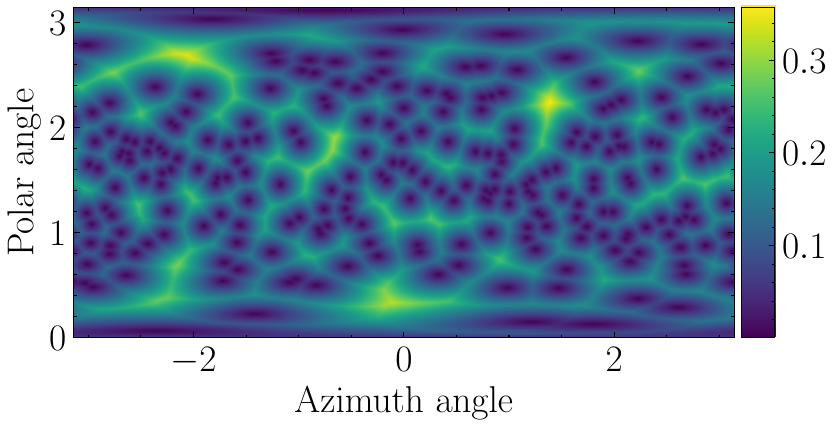}
	\caption{Spherical distance transform of the point cloud in Fig.~\ref{fig:library_test_example}(a). The local maximums correspond to centers of sparse regions where few points are present and the local minimums corresponds to points of $\pcx$.}
	\label{fig:spherical_distance_transform}
\end{figure}

We discretized the unit sphere into $D\times D$ pixels and computed the spherical distance transform of $\pcx$, which is a $D\times D$ spherical image. For pixel $(p,q)$, the pixel value $d_{pq}$ is the minimum distance between its center $\vecr_{pq}$ and all points in $\pcx$:
\begin{equation}
    d_{pq}=d(\vecr_{pq},\pcx)=\underset{k}{\min}~\|\vecx_k-\vecr_{pq}\|
\end{equation}
In order to solve Eq.~\eqref{eq:opt_1}, we locate the pixel $(p^*,q^*)$ that $\rotation\vecy_j$ falls into and approximate the minimum distance $d(\rotation\vecy_j,\pcx)$ by $d_{p^*q^*}$. In our study, we set $D=400$, corresponding to a pixel size of \SI{15.71}{\milli\radian}$\times$\SI{7.85}{\milli\radian}, which gave us sufficient accuracy for point cloud registration. Figure~\ref{fig:spherical_distance_transform} shows the distance transform of the point cloud in Fig.~\ref{fig:library_test_example}(a).

\subsubsection{Solving Eq.~\eqref{eq:opt_2}{ to find the globally minimal cost}}
Eq.~\eqref{eq:opt_2} evaluates the metric function $d_2(\pcx,\pcy)$ by finding the global minimum of the cost function $J$. Gradient-based optimization approaches are not applicable to this problem due to the non-convexity of $J$, which tends to be trapped at local minimums~\cite{jain2017non}. We adopted the Globally Optimal Iterative Closest Point (Go-ICP) algorithm~\cite{yang2013go,yang2015go} to perform the global minimization, with the simplification that only rotation transformation needs to be considered. The simplified Go-ICP algorithm is outlined in Algorithm~\ref{alg:go-icp}. In brief, we divide the search space into smaller subspaces and calculate the lower-bound and upper-bound of the global minimum in each subspace; subspaces with lower bounds higher than the current minimum are pruned. Our implementation used the lower bound and upper bound derived in ~\cite{yang2015go}. The process is repeated until the desired accuracy or maximum number of iterations is achieved. 

\begin{figure}[!htbp]
	\centering
	\includegraphics[width=.5\linewidth]{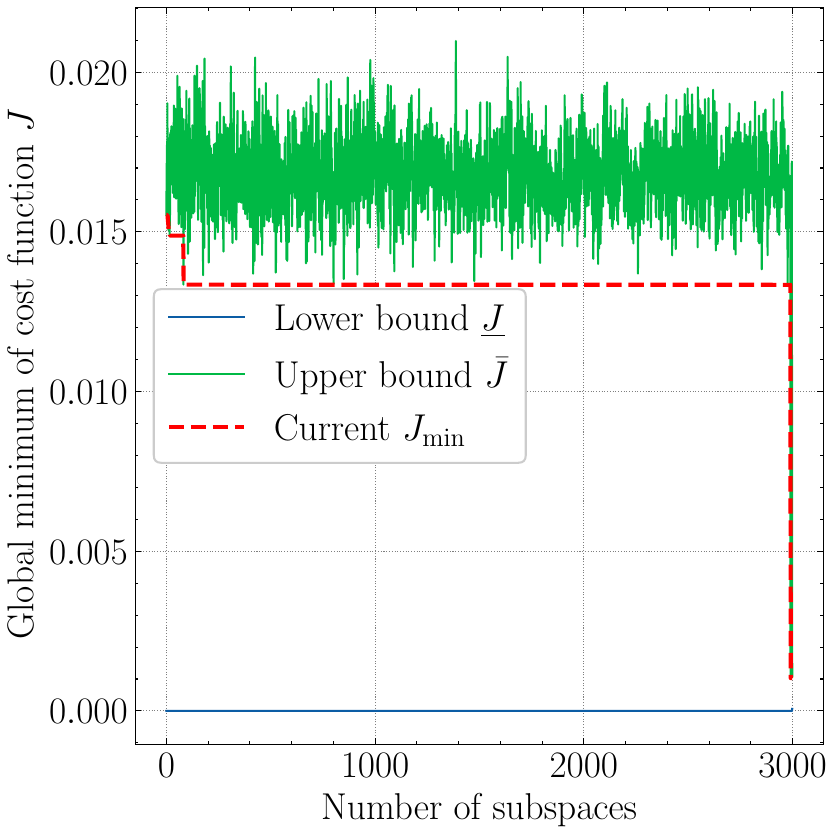}
	\caption{Lower bounds and upper bounds of $J_{\min}$ in each subspace and the evolution of current best estimation $J_{\min}$. The convergence criteria is that the difference between the current $J_{\min}$ and lower bound $\underline{J}$ must be below $\epsilon=0.001$. Convergence was reached after exploring 3000 subspaces (441 iterations).}
	\label{fig:goicp_bounds_evolution}
\end{figure}

\begin{figure}[!htbp]
	\centering
	\includegraphics[width=\linewidth]{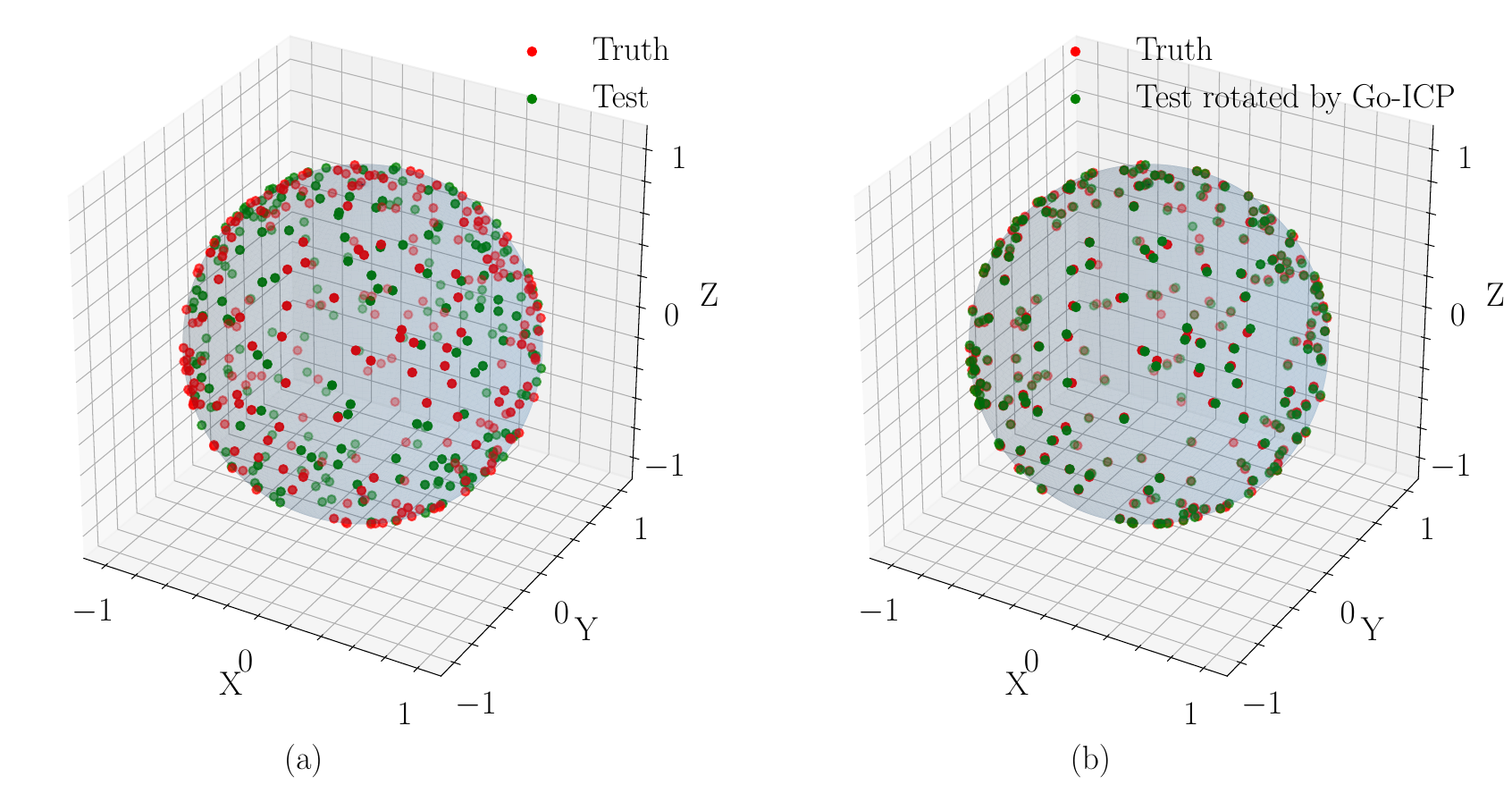}
	\caption{(a) A point cloud from the library (truth) and a point cloud from test dataset (test). (b) Apply the optimal rotation found by Go-ICP to the test point cloud.}
	\label{fig:goicp_ex}
\end{figure}

As an example, we applied Go-ICP to align the two point clouds in Fig.~\ref{fig:library_test_example}(a) and Fig.~\ref{fig:library_test_example}(c). Fig.~\ref{fig:goicp_bounds_evolution} shows the evolution of estimation of $J_{\min}$ as the subspaces gets smaller as well as the lower bound $\underline{J}$ and upper bound $\bar{J}$ of each subspace. Convergence is reached when the difference between current $J_{\min}$ and lower bound $\underline{J}$ is below $\epsilon=0.001$. Figure~\ref{fig:goicp_ex} shows that after applying Go-ICP, the two point clouds are well-aligned, resulting in a small $d_2$.
\begin{algorithm}[H]
    \caption{Evaluation of metric $d_2(\pcx, \pcy)$ using Go-ICP}\label{alg:go-icp}
    \begin{algorithmic}[1]
        \Procedure{Rotation-only Go-ICP}{}\newline
        \textbf{Input:} Point cloud $\pcx$. Point cloud $\pcy$. Threshold $\epsilon$. Rotation search space $C_0$.\newline
        \textbf{Initialize:} Priority queue $Q$, $J_{\min}=\infty$.
        \State{Calculate the lower bound of $J(\rotation,\pcx,\pcy)$ for $\rotation\in C_0$.}
        \State{Append $C$ to $Q$.}
        \While{$Q$ is not empty}
                \State Read $C$ from $Q$ with the lowest lower-bound $\underline{J}$.
                \State If $|J_{\min}-\underline{J}|<\epsilon$, break the loop
                \State Divide $C$ into 8 subspaces.
                \For{each subspace $C_i$}
                    \State{Calculate the lower bound $\underline{J}$}
                    \State{If $\underline{J}\geq J_{\min}$, process the next subspace}
                    \State{Calculate the upper bound $\bar{J}$}
                    \State{If $\bar{J}<J_{\min}, J_{\min}=\bar{J}$}
                    \State{Add $C_i$ to $Q$}
              \EndFor
        \EndWhile
        \State {\textbf{Return:} $J_{\min}$}
        \EndProcedure
    \end{algorithmic}
\end{algorithm}


\subsection{Generation of Library Dataset and Test Dataset}
In this section, we describe the methods used to generate a library dataset of 100,000 random and unique pebbles, as well as a test dataset of 10,000 pebbles for testing the identification algorithm. A TRISO-fueled pebble has a diameter of \SI{60}{\mm} and contains approximately 10,000 TRISO-fuel particles, which are randomly and uniformly distributed within a fuel zone of \SI{25}{\mm} radius~\cite{zhu2018uniformity}. The diameter of the coated fuel particle is \SI{910}{\um} by design~\cite{terry2005evaluation}, which is the minimum distance between two kernels. 

\begin{figure}[!htbp]
	\centering
	\includegraphics[width=.5\linewidth]{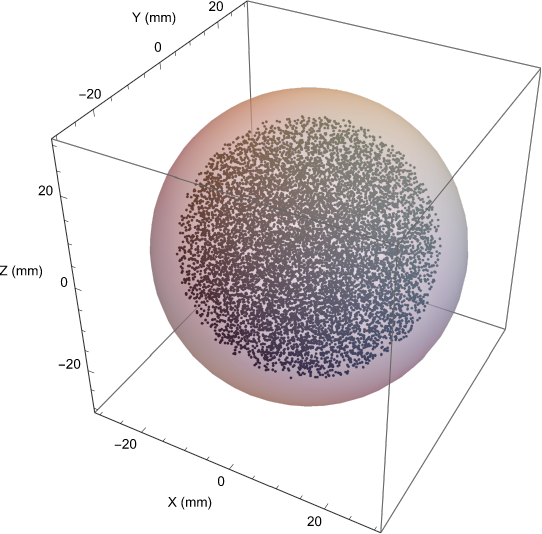}
	\caption{A random pebble containing 10254 randomly-distributed particles within a \SI{25}{\mm} radius fuel zone.}
	\label{fig:pebble_example}
\end{figure}

The library dataset was generated using rejection sampling. We first generated 45,000 random points uniformly distributed within a cube of \SI{60}{\mm} side length centered at the origin. Then we rejected the points with distance to origin above \SI{30}{\mm}. Finally, we rejected the points with distance to their nearest neighbor less than \SI{910}{\um}. The number of remaining points is approximately 10,000-11,000, which forms a random pebble. We repeated this procedure for 100,000 times to generate the library dataset but with a unique seed to initiate the random number generation each time to ensure uniqueness of each pebble. Figure~\ref{fig:pebble_example} shows a random pebble generated using this procedure. 

The test dataset was generated by applying random rotations and noises to the library dataset. For a point cloud of $n$ points, we first applied a random rotation to all the points. Then we added Gaussian noises of mean 0 and standard deviation $\sigma$ to each point, where $\sigma$ represents the noise level associated with measured kernel positions. Finally we randomly removed $n_1$ points to simulate missing fuel kernels (false negatives) and inserted $n_2$ random points to simulate non-existing fuel particles (false positives) introduced during image segmentation. $n_1$ and $n_2$ were sampled from the range $[0,np/2]$ and $p$ represents the maximum fraction of outliers (including missing kernels and non-existing kernels) due to segmentation inaccuracy. A total of nine test datasets were generated, with $\sigma=0.5, 1.0,$ and \SI{1.5}{\mm} and  $p=10\%,20\%,30\%$. We examined if the proposed algorithm can correctly identify the tested pebble at various noise levels $\sigma$ and outlier fractions $p$. 

\section{Results}\label{sec:results}
\subsection{Experimental Results}
{This section reports the results on the extraction of kernel distribution through X-ray CT. Overall, we achieved a maximum outlier fraction of $10\%$ and a maximum positional error of \SI{500}{\um}, and the processing time was approximately \SI{40}{s}, compatible with reaction operation.}
\subsubsection{X-ray Projection}
\begin{figure}[!htbp]
	\centering
	\includegraphics[width=.5\linewidth]{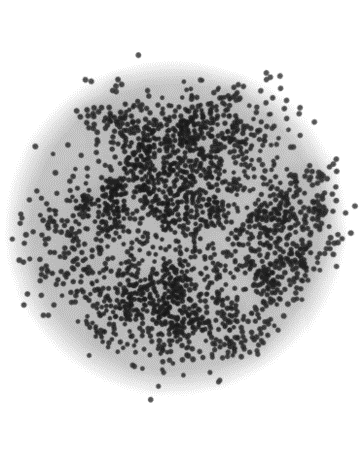}
	\caption{Projection at scanning angle = $0^{\circ}$ of mock-up fuel sample. The dark color indicates high X-ray attenuation.}
	\label{fig:xray-projection}
\end{figure}

Figure~\ref{fig:xray-projection} shows the projection of the WC-loading sample at projection angle = 0$^{\circ}$. The high-density WC kernels shown as black dots can be clearly discriminated from the low-density Lucite matrix.

\subsubsection{Image Reconstruction}
\begin{figure}[!htbp]
    \begin{subfigure}[t]{0.51\linewidth}
        \centering
        \includegraphics[width=.95\linewidth]{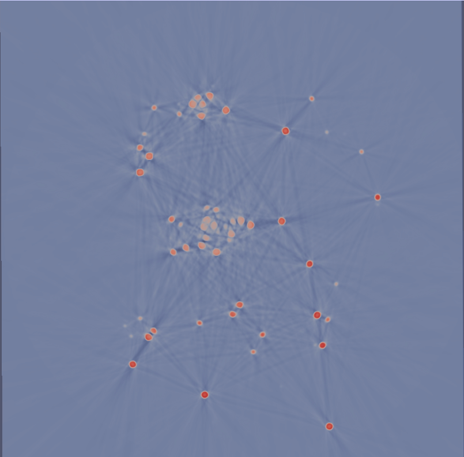}
        \caption{NSI reconstruction}
    \end{subfigure}\hfil
    \begin{subfigure}[t]{0.49\linewidth}
        \centering
        \includegraphics[width=.95\linewidth]{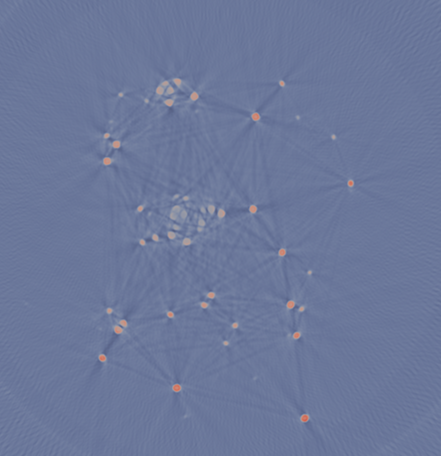}
        \caption{ASTRA reconstruction}
    \end{subfigure}
	\caption{Comparison of images reconstructed using NSI's proprietary software and the ASTRA toolbox.}
  \label{fig:WC_1p_astra_recon}
\end{figure}

In the ASTRA reconstruction, we downsampled the projections from $1536\times1920$ pixels to $768\times960$ pixels and used only 360 scans {out of 2520 scans} to accelerate the computation. Figure~\ref{fig:WC_1p_astra_recon} shows the comparison of images of the central slice reconstructed using the NSI's reconstruction software and ASTRA. The high-density WC kernels can easily be discriminated from the background, and the reconstruction qualities are comparable.

\subsubsection{Image Segmentation}
\begin{figure}[!htbp]
	\centering
	\includegraphics[width=\linewidth]{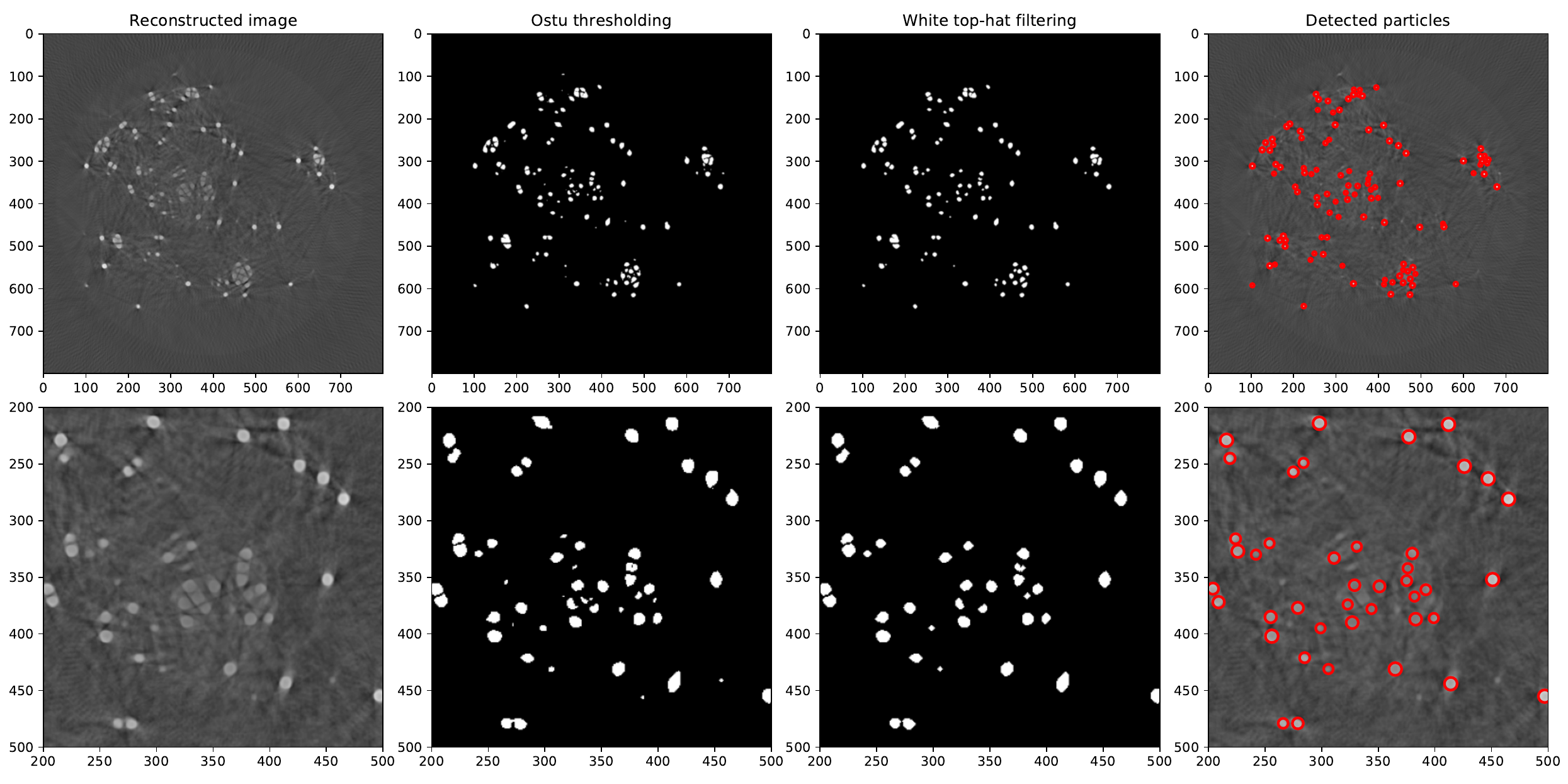}
	\caption{Steps to segment kernels from the reconstructed image. The first column shows the image of a slice of the sample reconstructed by ASTRA; the second column shows the binary image obtained by applying Ostu's thresholding; the third column shows the binary image after applying the white top-hat filter to remove small regions; the last column shows the identified kernels overlaid on the reconstructed image. The bottom row is the zoomed-in view of the top row within the [200, 500]$\times$[200, 500] region.}
	\label{fig:segmentation_result}
\end{figure}

{The kernels were shown as bright regions on the reconstructed image. In this step we extracted the kernels by segmenting the reconstructed image. }Figure~\ref{fig:segmentation_result} shows the segmentation process for a 2D slice of the 3D reconstruction.{ A binary image was obtained by applying an intensity threshold found using Ostu's method (second column). The reconstruction artifacts shown as small white regions were removed through the white top-hat filter (third column). Finally, cross sections of kernels shown as small disks were identified and their centroids were calculated (fourth column).} The algorithm is able to segment the kernels correctly in most cases.

\begin{figure}[!htbp]
	\centering
	\includegraphics[width=\linewidth]{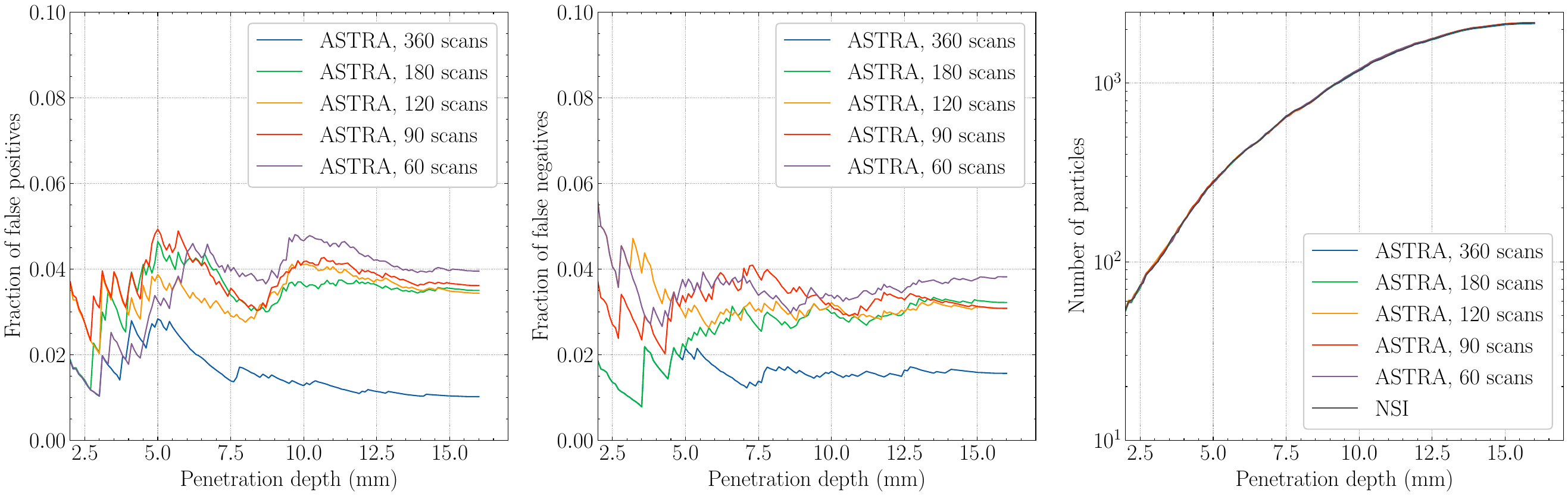}
	\caption{False positive rate and false negative rate as a function of the penetration depth and the total number of scans.}
	\label{fig:segmentation_FPFN}
\end{figure}

{We then calculated the false positive and false negative rates by comparing the segmentation based on ASTRA and NSI reconstructions. These two rates quantify the inaccuracy in the extracted kernel distribution, based on which test datasets can be generated. }Figure~\ref{fig:segmentation_FPFN} shows the false positive rate and false negative rate as a function of the penetration depth and the total number of scans. Lower false positive and false negative rates mean more accurate segmentation. The outlier fraction, defined as the sum of false positive and false negative rates, is approximately 4\% when there are 360 scans and 10\% for the other cases using fewer scans. The false positive and negative rates do not increase significantly when we reduce the number of scans from 360 to 60, which means we can potentially reduce the CT scan time to \SI{16}{\s} and reconstruction time to \SI{3.2}{\s} based on Table~\ref{table:ct_seg_time}. The reconstruction was performed on a NVIDIA Quadro P4000 GPU and the segmentation was performed on an i9-7920X CPU @ 2.90GHz with 24 threads in parallel. 

\begin{table}[!htbp]
\centering
\caption{CT acquisition and image processing time for different number of scans. The acquisition time is estimated based on X5000's scanning rate, which is 30fps at $768\times960$ resolution.}\label{table:ct_seg_time}
\begin{tabular}{|c|c|c|c|c|}
\hline
Scans & Acquisition (s) & Reconstruction (s) & Segmentation (s)    & Total (s) \\ \hline
360             & 97              & 9             & \multirow{5}{*}{22} & 128  \\ \cline{1-3} \cline{5-5} 
180             & 49              & 5             &                     & 76    \\ \cline{1-3} \cline{5-5} 
120             & 33              & 4              &                     & 59    \\ \cline{1-3} \cline{5-5} 
90              & 24              & 4              &                     & 50    \\ \cline{1-3} \cline{5-5} 
60              & 16              & 3              &                     & 41    \\ \hline
\end{tabular}
\end{table}

\subsection{Computational Results}
{In this section we report the results of identifying pebbles with positional noises and outliers from a library of 100,000 pebbles. The identification accuracy was 100\% for noise level below \SI{1.5}{\mm} and outlier fraction below 30\%. The average identification time for a single pebble was approximately \SI{6.6}{\s}.}
\subsubsection{Search Space Reduction Based on Rotation-invariant Feature Matching}
\begin{figure}[!htbp]
    \begin{subfigure}[t]{0.33\linewidth}
        \centering
        \includegraphics[width=\linewidth]{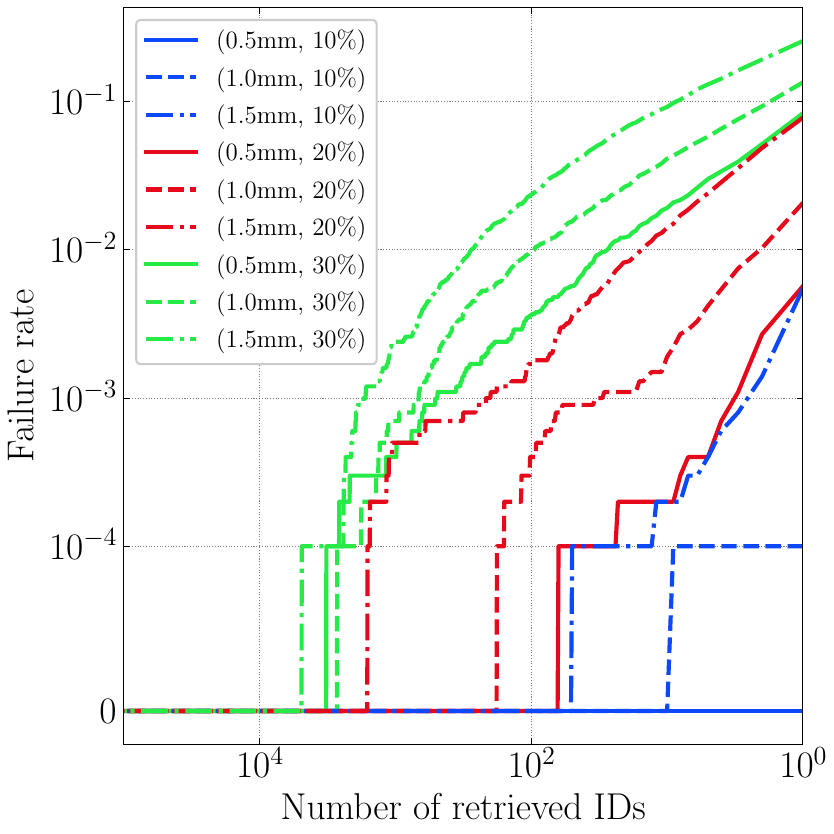}
        \caption{Single shell}
        \label{fig:failure_rate_shell1}
    \end{subfigure}\hfil
    \begin{subfigure}[t]{0.33\linewidth}
        \centering
        \includegraphics[width=\linewidth]{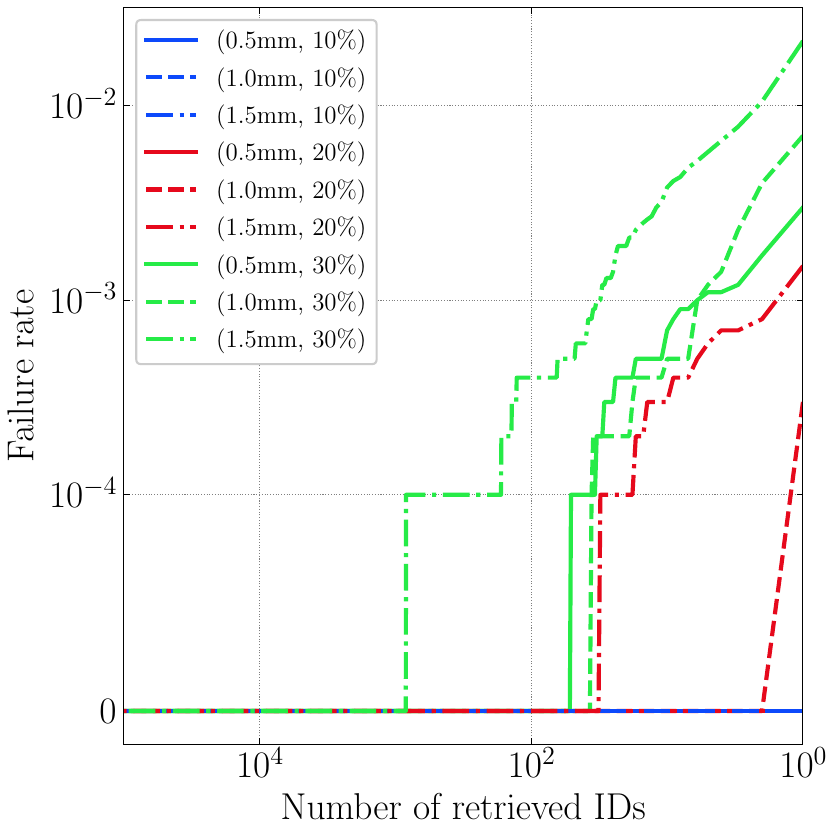}
        \caption{Two shells}
        \label{fig:failure_rate_joint}
    \end{subfigure}\hfil
    \begin{subfigure}[t]{0.33\linewidth}
        \centering
        \includegraphics[width=\linewidth]{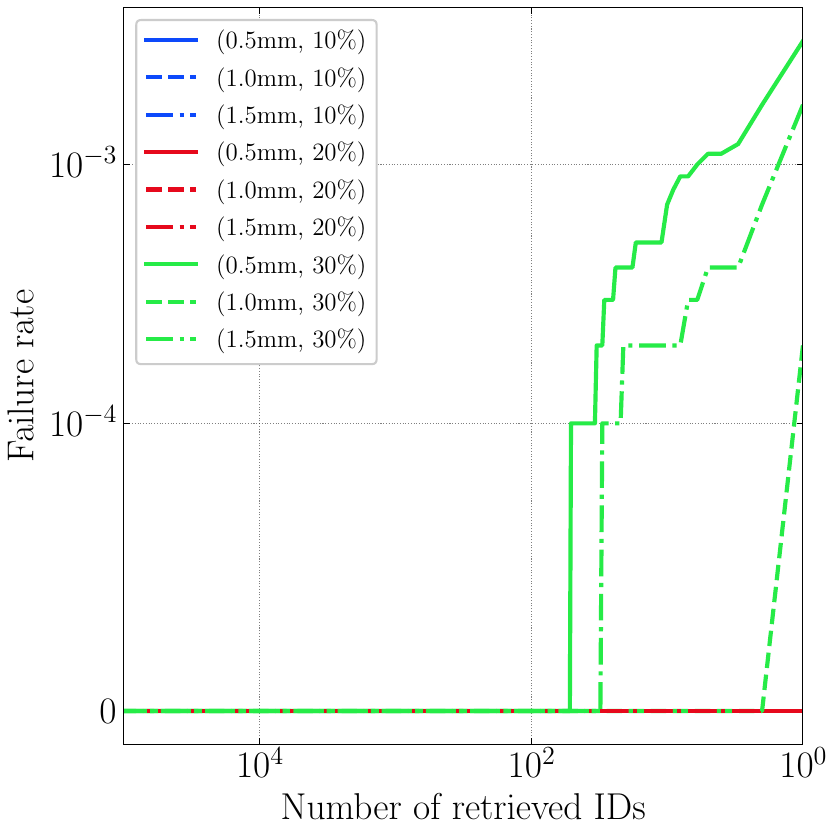}
        \caption{Three shells}
        \label{fig:failure_rate_joint_3shells}
    \end{subfigure}
	\caption{Test failure rate as a function of number of returned pebbles at different noise levels and outlier fractions, when (a) a single spherical shell, (b) two shells, and (c) three shells were used for identification. A test case where the true pebble ID is not found within the list returned by the algorithm is called a failure.}
  \label{fig:failure_rate_vs_noises}
\end{figure}

First, we examined the performance of the search-space reduction algorithm at different noise levels ($\sigma$=0.5,1,\SI{1.5}{\mm}) and outlier fractions ($p$=10\%, 20\%, 30\%). Given a test pebble, the search-space reduction algorithm should traverse the library of 100,000 pebbles and return a short list of pebbles in which there is one matching the input. If the true ID of the test pebble is not within the returned list, we call it a test failure. For each combination of $\sigma$ and $p$, we calculated the test failure rate as a function of the number of returned pebbles, as shown in Fig.~\ref{fig:failure_rate_shell1}. When $\sigma\leq$ \SI{0.5}{mm} (diameter of the fuel kernel) and $p\leq10\%$, the failure rate is 0, indicating that the true pebble ID is always the first element of the returned list. As $\sigma$ or $p$ increases, the algorithm's performance degrades and thousands of IDs need to be retrieved to ensure the true pebble ID is not missed (zero failure rate). These results were achieved using the outermost 2\%-3\% points with \SI{24.8}{\mm}$<r<$\SI{25}{\mm}. To improve identification accuracy, we extracted an independent rotation-invariant feature from the spherical shell with \SI{24.6}{\mm}$<r<$\SI{24.8}{\mm} and used both features for identification. The identification results using two and three shells are shown in Fig.~\ref{fig:failure_rate_joint},~\ref{fig:failure_rate_joint_3shells}, from which we observe that the failure rate is significantly reduced for all test datasets. We achieved a zero failure rate using the top-100 IDs and the three outermost shells. If more shells were to be used, the failure rate would be further reduced. 
The average computational time per test case was \SI{5.2}{\ms}, \SI{0.24}{s}, and \SI{0.37}{s} using one shell, two shells, and three shells, which is compatible with the online-refueling scheme. The tests were performed on an i9-7920X CPU @ 2.90GHz with no parallelization.

\subsubsection{Pebble Identification Based on Point Cloud Registration}

\begin{figure}[!htbp]
    \begin{subfigure}[t]{0.5\linewidth}
        \centering
        \includegraphics[width=\linewidth]{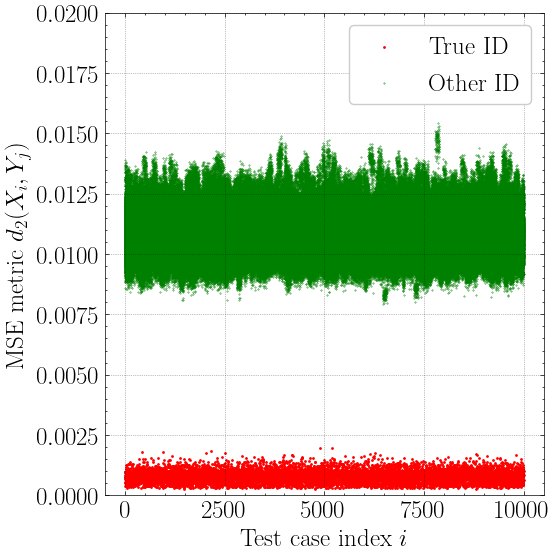}
        \caption{$\sigma=$\SI{0.5}{\mm},$p=10\%$}
        \label{fig:d2_metrics_0p5_10p}
    \end{subfigure}\hfil
    \begin{subfigure}[t]{0.5\linewidth}
        \centering
        \includegraphics[width=\linewidth]{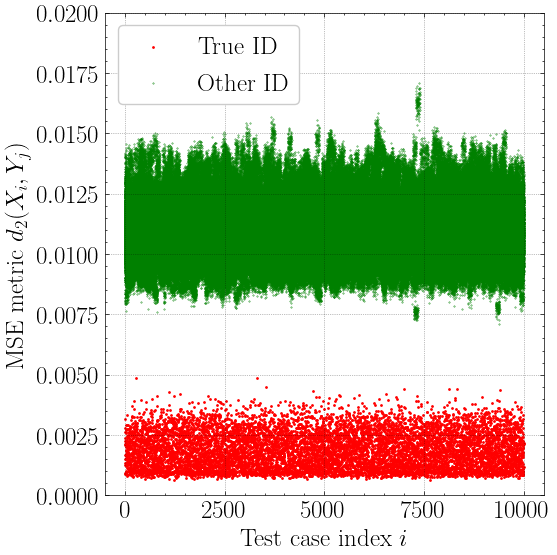}
        \caption{$\sigma=$\SI{1.5}{\mm},$p=30\%$}
        \label{fig:d2_metrics_1p5_30p}
    \end{subfigure}
	\caption{MSE calculated using Go-ICP for true pebble ID and other IDs.}
  \label{fig:d2_metrics}
\end{figure}

We tested the Go-ICP based identification algorithm on two test datasets with $(\sigma=$\SI{0.5}{\mm} and $p=10\%)$ and $(\sigma=$\SI{1.5}{\mm} and $p=30\%)$. For each test case $i$, we constructed a new library $\{\pcy_j|j\in \mathcal{J}_i\}$, where $\mathcal{J}_i$ represents the collection of 100 pebble IDs retrieved in the previous step. We then applied the Go-ICP algorithm to calculate the metric $d_2(\pcx_i, \pcy_j)$ between the input pebble $\pcx_i$ and each library pebble $\pcy_j$, which is defined as the minimum MSE over all rotations. The convergence threshold $\epsilon$ of Go-CIP was set to 0.001 and the maximum number of iterations was set to 5000 to ensure convergence. Figure~\ref{fig:d2_metrics} shows the distribution of $d_2(\pcx_i, \pcy_j)${. The x-axis is the index of the test pebble ranging from 1 to 10,000, and the y-axis is the error after alignment}. The red data points represent the alignment error when $j$ corresponds to the true pebble ID of $\pcx_i$ and the green data points for the other 99 IDs. We observed an excellent separation between the two sets of IDs, enabling ID retrieval by thresholding $d_2$. For both test datasets, we applied a threshold of 0.006, and the pebble IDs were correctly retrieved for all of the test cases, achieving a failure rate of 0. Computational time per test case is \SI{6.6}{\s} on average, in parallel with 10 cores.

\section{Discussion and Conclusions}\label{sec:conclusion}
In this work, we developed a procedure to extract the spatial distribution of fuel kernels in a TRISO-fueled pebble using X-ray CT and experimentally demonstrated it on a mock-up fuel sample with X-ray attenuation close to the attenuation of an actual fuel pebble.
The present study is limited to the analysis of X-ray CT images. The developed algorithms were demonstrated on a large data set of 100,000 simulated pebbles, and kernel identification performances were determined experimentally using an industrial X-ray CT scanner and mock-up fuel compact, with an X-ray attenuation comparable to the attenuation of an actual TRISO-fueled pebble.
For spent TRISO-fueled pebbles, the photons emitted by the spent fuel may saturate the X-ray detector~\cite{sawicka1990computed}. In this case, neutron tomography~\cite{LEHMANN2003745,ZHANG2023101434} or other methods that allow for reliable extraction of the positions of outermost kernels can be applied. The computational method described in this work also applies to other identification concepts based on the spatial distribution of external identifiers such as ZrO\textsubscript{2}~\cite{gitau2012development,gariazzo2021nuclear}.
Our study focuses on the identification of TRISO-fueled pebbles of spherical shape, but the approach can be easily generalized to fuel elements of arbitrary shapes {as long as there is a random 3D distribution of kernels in the fuel.} In fact, spherical pebbles are the most difficult to identify of all because there are three degrees of freedoms (rotation angle and axis) to be optimized. For cylindrical fuel elements, the rotation axis is known and there is only one degree of freedom (angle), and faster 1-D registration method can be applied~\cite{cai2019practical}.

{The demonstration on WC-loaded sample allowed us to estimate the error of the extracted kernel distribution and the total time needed for extraction. }Our experimental results show that the error associated with the kernel position is less than \SI{500}{\um} and the percentage of mis-identified particles is below 10\%. { The total imaging processing time required to extract the kernel distribution was approximately \SI{40}{\s}.} It should be noted that the kernel distribution in our sample is not uniform, leading to difficulty in the segmentation of densely-distributed WC kernels. We expect a higher segmentation accuracy can be achieved for actual TRISO-fueled pebbles, where the kernel distribution is sparser~\cite{yu20173d,helmreich2020new}.
Additionally, a detailed image of the deeper region of the pebble is not necessary since our identification method only relies on the outermost kernels that are the easiest to capture by X-ray CT, therefore further relaxing the constraints of the CT scan.

We have developed a coarse-to-fine strategy to efficiently and accurately identify a TRISO-fueled pebble from a library of 100,000 pebbles based on their unique kernel distribution. The coarse-search step effectively reduces the size of search space by 99.9\%, by comparing the rotation-invariant features of the kernel distribution. The fine-search step compares each candidate in the size-reduced library to the pebble under inspection and eventually returns the one that aligns best with the input pebble. Our numerical experiments showed that this approach achieves 100\% identification accuracy with positional noise up to \SI{1.5}{\mm} and outlier fraction up to 30\%. The proposed pebble identification algorithm allows us to obtain not only the pebble ID but also the 3D rotation that the pebble has gone through. The latter can be useful for tracking the movements of individual TRISO fuel particles during the fuel's lifespan.

\begin{figure}[!htbp]
	\centering
	\includegraphics[width=\linewidth]{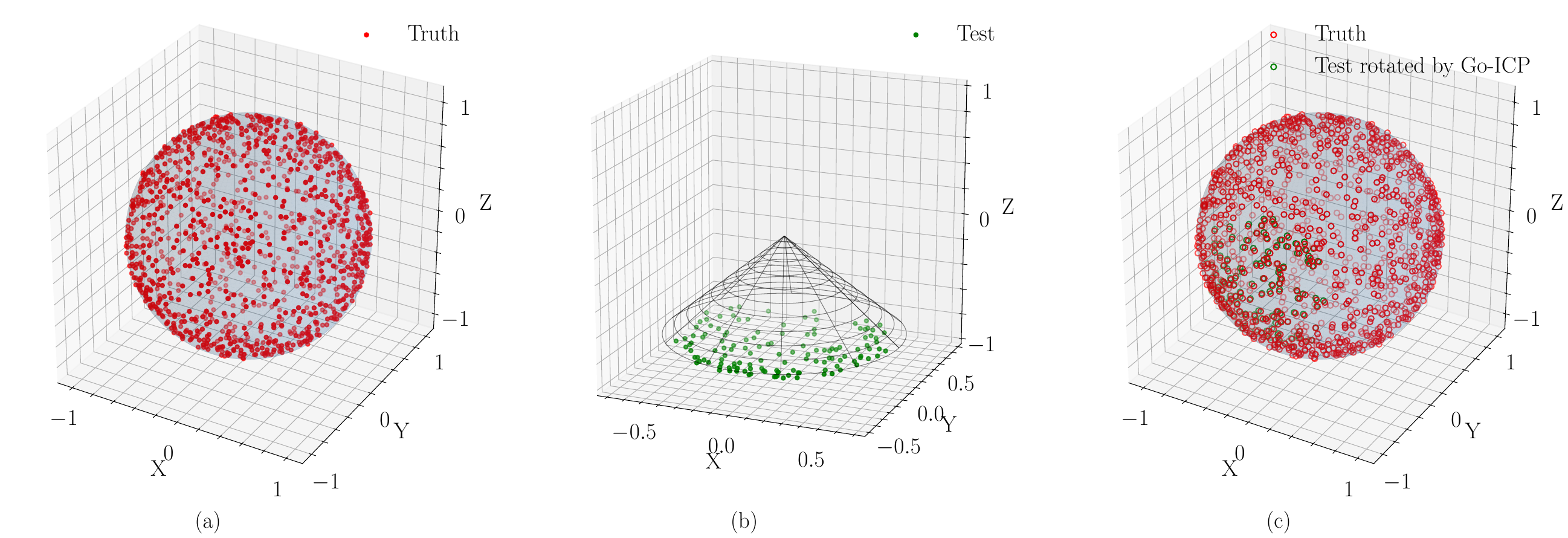}
	\caption{(a) An intact pebble (truth). (b) Apply a random rotation and add noises of $\sigma=$\SI{0.5}{\mm} to the truth and remove 90\% particles to simulate a damaged pebble (test). (c) Apply the optimal rotation found by Go-ICP to the test.}
	\label{fig:partial_pebble_alignment}
\end{figure}

The rotation invariant feature is extracted from the kernel distribution in a spherical shell, which implicitly requires the pebble to be intact when it exits the PBR's core. However, in some circumstances, the pebble may be damaged during fuel circulation. The coarse search algorithm no longer applies due to the large outlier fraction, while the fine search algorithm based on Go-ICP, although time-consuming, is shown to be robust against large outlier fractions. Figure~\ref{fig:partial_pebble_alignment} shows the alignment of a damaged pebble with 90\% kernels missing with the ground truth in the library, calculated using Go-ICP. In practice, the damaged pebbles would result in a non-identified pebble, which will be collected separately into waste containers and the identification can therefore be performed offline using Go-ICP. 

The time to identify a single fuel pebble from 100,000 pebbles is less than \SI{7}{\s} in parallel with 10 cores. If needed, the computational time can be further reduced by extracting more rotation-invariant features in the coarse-search step to reduce the number of retrieved candidates. Additionally, other independently-measured inherent signatures associated with the pebble, such as the fuel burnup or residual \textsuperscript{235}U mass obtained through gamma-ray spectroscopy~\cite{SU2006686} or neutron multiplicity counting~\cite{fang2023boron,fang2023feasibility}, can be relied upon to narrow the search. For example, pebbles in the library with higher burnup or lower \textsuperscript{235}U mass than the inspected pebble's can be immediately rejected. 

In conclusion, we have developed a coarse-to-fine approach to efficiently and accurately identify a TRISO-fueled pebble exiting a PBR core. The identification relies on the unique spatial distribution of fuel kernels associated with each pebble and we demonstrated experimentally that this distribution can be accurately extracted through X-ray CT in \SI{41}{\s}, including measurement and data processing time, with a maximum positional error of \SI{0.5}{\mm} and outlier fraction of 10\%. The coarse-to-fine approach allows the retrieval of the ID of an unknown pebble from a library of 100,000 pebbles in \SI{7}{\s}.
The identification accuracy of our method is 100\% in 10,000 tests with measurement error up to \SI{1.5}{\mm} and outlier fraction up to 30\%. The proposed approach will be beneficial for fuel management and safeguarding SNM in PBRs.

\section*{ACKNOWLEDGMENTS}
This work was funded in part by STTR-DOE grant DE-SC0020733. We would also like to thank BWXT for providing the WC-loaded sample.

\bibliographystyle{elsarticle-num} 
\bibliography{references}

\begin{thebibliography}{10}
\expandafter\ifx\csname url\endcsname\relax
  \def\url#1{\texttt{#1}}\fi
\expandafter\ifx\csname urlprefix\endcsname\relax\def\urlprefix{URL }\fi
\expandafter\ifx\csname href\endcsname\relax
  \def\href#1#2{#2} \def\path#1{#1}\fi

\bibitem{boll2023advanced}
A.~M. Boll, W.~E. Windes, Advanced reactors development in usa, Tech. rep.,
  Idaho National Laboratory (INL), Idaho Falls, ID (United States) (2023).

\bibitem{kelly2014generation}
J.~E. Kelly, Generation iv international forum: A decade of progress through
  international cooperation, Progress in Nuclear Energy 77 (2014) 240--246.

\bibitem{lorusso2018gen}
P.~Lorusso, S.~Bassini, A.~Del~Nevo, I.~Di~Piazza, F.~Giannetti, M.~Tarantino,
  M.~Utili, Gen-iv lfr development: status \& perspectives, Progress in Nuclear
  Energy 105 (2018) 318--331.

\bibitem{pioro2022handbook}
I.~Pioro, Handbook of Generation IV Nuclear Reactors: A Guidebook, Woodhead
  Publishing, 2022.

\bibitem{kadak2005future}
A.~C. Kadak, A future for nuclear energy: pebble bed reactors, International
  journal of critical infrastructures 1~(4) (2005) 330--345.
\newblock \href {https://doi.org/10.1016/j.pnucene.2022.104175}
  {\path{doi:10.1016/j.pnucene.2022.104175}}.

\bibitem{zhang2016shandong}
Z.~Zhang, Y.~Dong, F.~Li, Z.~Zhang, H.~Wang, X.~Huang, H.~Li, B.~Liu, X.~Wu,
  H.~Wang, et~al., The shandong shidao bay 200 mwe high-temperature gas-cooled
  reactor pebble-bed module (htr-pm) demonstration power plant: an engineering
  and technological innovation, Engineering 2~(1) (2016) 112--118.

\bibitem{mulder2021x}
E.~J. Mulder, X-energy’s xe-100 reactor design status, Presentation to
  National Academy of Sciences, May 26 (2021).

\bibitem{international2010high}
I.~A.~E. Agency, High temperature gas cooled reactor fuels and materials, IAEA,
  2010.

\bibitem{topan2016study}
S.~Topan, I.~Dwi, et~al., Study on fuel multipass effect on core performance of
  small pebble bed reactor (2016).

\bibitem{forsberg2009safeguards}
C.~W. Forsberg, D.~L. Moses, Safeguards challenges for pebble-bed reactors
  designed by people’s republic of china, Global Nuclear Security Technology
  Division, http://www. osti. gov/bridge (2009).

\bibitem{kovacic2020advanced}
D.~Kovacic, P.~Gibbs, L.~Worrall, R.~Hunneke, J.~Harp, J.~Hu, Advanced reactor
  safeguards: Nuclear material control and accounting for pebble bed reactors,
  ORNL 1849 (2020).

\bibitem{durst2009nuclear}
P.~C. Durst, D.~Beddingfield, B.~Boyer, R.~Bean, M.~Collins, M.~Ehinger,
  D.~Hanks, D.~L. Moses, L.~Refalo, Nuclear safeguards considerations for the
  pebble bed modular reactor (pbmr), Tech. rep., Idaho National Lab.(INL),
  Idaho Falls, ID (United States) (2009).

\bibitem{haire2007tags}
M.~J. Haire, Tags to track illicit uranium and plutonium, Tech. rep., Oak Ridge
  National Lab.(ORNL), Oak Ridge, TN (United States) (2007).

\bibitem{gitau2012development}
E.~T.~N. Gitau, Development and evaluation of a safeguards system concept for a
  pebble-fueled high temperature gas-cooled reactor, Ph.D. thesis (2012).

\bibitem{helmreich2020new}
G.~W. Helmreich, J.~D. Hunn, D.~R. Brown, B.~J. Blamer, New method for analysis
  of x-ray computed tomography scans of triso fuel forms, Nuclear Engineering
  and Design 357 (2020) 110418.

\bibitem{vrinda2019triso}
K.~Vrinda~Devi, J.~Dubey, J.~Gupta, I.~Shaikh, Triso fuel volume fraction and
  homogeneity: a nondestructive characterization, Nuclear Science and
  Techniques 30~(3) (2019) 1--7.

\bibitem{kane20223d}
J.~J. Kane, D.~W. Marshall, N.~L. Cordes, W.~C. Chuirazzi, B.~Kombaiah, I.~van
  Rooyen, J.~D. Stempien, 3d analysis of triso fuel compacts via x-ray computed
  tomography, Journal of Nuclear Materials 565 (2022) 153745.

\bibitem{yu20173d}
G.~Yu, Y.~Du, X.~Xiang, Y.~Liu, Z.~Li, X.~Wang, et~al., 3d nondestructive
  visualization and evaluation of triso particles distribution in htgr fuel
  pebbles using cone-beam computed tomography, Science and Technology of
  Nuclear Installations 2017 (2017).

\bibitem{KWAPIS2021103913}
E.~H. Kwapis, H.~Liu, K.~C. Hartig,
  \href{https://www.sciencedirect.com/science/article/pii/S0149197021002742}{Tracking
  of individual triso-fueled pebbles through the application of x-ray imaging
  with deep metric learning}, Progress in Nuclear Energy 140 (2021) 103913.
\newblock \href {https://doi.org/https://doi.org/10.1016/j.pnucene.2021.103913}
  {\path{doi:https://doi.org/10.1016/j.pnucene.2021.103913}}.
\newline\urlprefix\url{https://www.sciencedirect.com/science/article/pii/S0149197021002742}

\bibitem{fang2022algorithms}
M.~Fang, A.~Di~Fulvio, Algorithms for triso fuel identification based on x-ray
  ct validated on tungsten-carbide compacts, arXiv preprint arXiv:2204.13774
  (2022).

\bibitem{terry2005evaluation}
W.~K. Terry, Evaluation of the initial critical configuration of the htr-10
  pebble-bed reactor, Tech. rep., Idaho National Lab.(INL), Idaho Falls, ID
  (United States) (2005).

\bibitem{osti_1419730}
C.~J. Werner, J.~S. Bull, C.~J. Solomon, F.~B. Brown, G.~W. McKinney, M.~E.
  Rising, D.~A. Dixon, R.~L. Martz, H.~G. Hughes, L.~J. Cox, A.~J. Zukaitis,
  J.~C. Armstrong, R.~A. Forster, L.~Casswell,
  \href{https://www.osti.gov/biblio/1419730}{Mcnp version 6.2 release notes} (2
  2018).
\newblock \href {https://doi.org/10.2172/1419730} {\path{doi:10.2172/1419730}}.
\newline\urlprefix\url{https://www.osti.gov/biblio/1419730}

\bibitem{Poludniowski2009}
G.~Poludniowski, G.~Landry, F.~DeBlois, P.~M. Evans, F.~Verhaegen,
  \href{https://dx.doi.org/10.1088/0031-9155/54/19/N01}{Spekcalc: a program to
  calculate photon spectra from tungsten anode x-ray tubes}, Physics in
  Medicine \& Biology 54~(19) (2009) N433.
\newblock \href {https://doi.org/10.1088/0031-9155/54/19/N01}
  {\path{doi:10.1088/0031-9155/54/19/N01}}.
\newline\urlprefix\url{https://dx.doi.org/10.1088/0031-9155/54/19/N01}

\bibitem{van2016fast}
W.~Van~Aarle, W.~J. Palenstijn, J.~Cant, E.~Janssens, F.~Bleichrodt,
  A.~Dabravolski, J.~De~Beenhouwer, K.~J. Batenburg, J.~Sijbers, Fast and
  flexible x-ray tomography using the astra toolbox, Optics express 24~(22)
  (2016) 25129--25147.

\bibitem{feldkamp1984practical}
L.~A. Feldkamp, L.~C. Davis, J.~W. Kress, Practical cone-beam algorithm, Josa a
  1~(6) (1984) 612--619.

\bibitem{otsu1979threshold}
N.~Otsu, A threshold selection method from gray-level histograms, IEEE
  transactions on systems, man, and cybernetics 9~(1) (1979) 62--66.

\bibitem{scikit-image}
S.~van~der Walt, J.~L. {S}ch\"onberger, J.~{Nunez-Iglesias}, F.~{B}oulogne,
  J.~D. {W}arner, N.~{Y}ager, E.~{G}ouillart, T.~{Y}u, the scikit-image
  contributors, \href{https://doi.org/10.7717/peerj.453}{scikit-image: image
  processing in {P}ython}, PeerJ 2 (2014) e453.
\newblock \href {https://doi.org/10.7717/peerj.453}
  {\path{doi:10.7717/peerj.453}}.
\newline\urlprefix\url{https://doi.org/10.7717/peerj.453}

\bibitem{kazhdan2003rotation}
M.~Kazhdan, T.~Funkhouser, S.~Rusinkiewicz, Rotation invariant spherical
  harmonic representation of 3 d shape descriptors, in: Symposium on geometry
  processing, Vol.~6, 2003, pp. 156--164.

\bibitem{mark_wieczorek_2019_3457861}
M.~Wieczorek, MMesch, E.~S. de~Andrade, I.~Oshchepkov, xoviat, B.~Xu,
  K.~Leinweber, A.~Walker,
  \href{https://doi.org/10.5281/zenodo.3457861}{Shtools/shtools: Version 4.5}
  (Sep. 2019).
\newblock \href {https://doi.org/10.5281/zenodo.3457861}
  {\path{doi:10.5281/zenodo.3457861}}.
\newline\urlprefix\url{https://doi.org/10.5281/zenodo.3457861}

\bibitem{wieczorek2018shtools}
M.~A. Wieczorek, M.~Meschede, Shtools: Tools for working with spherical
  harmonics, Geochemistry, Geophysics, Geosystems 19~(8) (2018) 2574--2592.

\bibitem{huang2021comprehensive}
X.~Huang, G.~Mei, J.~Zhang, R.~Abbas, A comprehensive survey on point cloud
  registration, arXiv preprint arXiv:2103.02690 (2021).

\bibitem{jain2017non}
P.~Jain, P.~Kar, et~al., Non-convex optimization for machine learning,
  Foundations and Trends{\textregistered} in Machine Learning 10~(3-4) (2017)
  142--363.

\bibitem{yang2013go}
J.~Yang, H.~Li, Y.~Jia, Go-icp: Solving 3d registration efficiently and
  globally optimally, in: Proceedings of the IEEE International Conference on
  Computer Vision, 2013, pp. 1457--1464.

\bibitem{yang2015go}
J.~Yang, H.~Li, D.~Campbell, Y.~Jia, Go-icp: A globally optimal solution to 3d
  icp point-set registration, IEEE transactions on pattern analysis and machine
  intelligence 38~(11) (2015) 2241--2254.

\bibitem{zhu2018uniformity}
L.~Zhu, X.~Xiang, Y.~Du, G.~Yu, Z.~Li, Y.~Peng, X.~Wang, et~al., Uniformity
  assessment of triso fuel particle distribution in spherical htgr fuel element
  using voronoi tessellation and delaunay triangulation, Science and Technology
  of Nuclear Installations 2018 (2018).

\bibitem{sawicka1990computed}
B.~Sawicka, R.~Murphy, G.~Tosello, P.~Reynolds, T.~Romaniszyn, Computed
  tomography of radioactive objects and materials, Nuclear Instruments and
  Methods in Physics Research Section A: Accelerators, Spectrometers, Detectors
  and Associated Equipment 299~(1-3) (1990) 468--479.

\bibitem{LEHMANN2003745}
E.~Lehmann, P.~Vontobel, A.~Hermann,
  \href{https://www.sciencedirect.com/science/article/pii/S0168900203023568}{Non-destructive
  analysis of nuclear fuel by means of thermal and cold neutrons}, Nuclear
  Instruments and Methods in Physics Research Section A: Accelerators,
  Spectrometers, Detectors and Associated Equipment 515~(3) (2003) 745--759.
\newblock \href {https://doi.org/https://doi.org/10.1016/j.nima.2003.07.059}
  {\path{doi:https://doi.org/10.1016/j.nima.2003.07.059}}.
\newline\urlprefix\url{https://www.sciencedirect.com/science/article/pii/S0168900203023568}

\bibitem{ZHANG2023101434}
Y.~Zhang, K.~G. Myhre, H.~Z. Bilheux, J.~A. Johnson, J.~C. Bilheux, C.~M.
  Parish, A.~J. Miskowiec, R.~D. Hunt, J.~Y. Lin,
  \href{https://www.sciencedirect.com/science/article/pii/S235217912300073X}{Non-destructive
  characterization of advanced nuclear fuel materials using neutron imaging},
  Nuclear Materials and Energy 35 (2023) 101434.
\newblock \href {https://doi.org/https://doi.org/10.1016/j.nme.2023.101434}
  {\path{doi:https://doi.org/10.1016/j.nme.2023.101434}}.
\newline\urlprefix\url{https://www.sciencedirect.com/science/article/pii/S235217912300073X}

\bibitem{gariazzo2021nuclear}
C.~Gariazzo, D.~B. Chojnowski, S.~Chirayath, Nuclear material control and
  accountancy approach for pebble fueled reactors using a novel pebble-type
  identification and classification technology, Tech. rep., Argonne National
  Lab.(ANL), Argonne, IL (United States) (2021).

\bibitem{cai2019practical}
Z.~Cai, T.-J. Chin, A.~P. Bustos, K.~Schindler, Practical optimal registration
  of terrestrial lidar scan pairs, ISPRS journal of photogrammetry and remote
  sensing 147 (2019) 118--131.

\bibitem{SU2006686}
B.~Su, Z.~Zhao, J.~Chen, A.~I. Hawari,
  \href{https://www.sciencedirect.com/science/article/pii/S0149197006000771}{Assessment
  of on-line burnup monitoring of pebble bed reactor fuel by passive neutron
  counting}, Progress in Nuclear Energy 48~(7) (2006) 686--702.
\newblock \href {https://doi.org/https://doi.org/10.1016/j.pnucene.2006.06.013}
  {\path{doi:https://doi.org/10.1016/j.pnucene.2006.06.013}}.
\newline\urlprefix\url{https://www.sciencedirect.com/science/article/pii/S0149197006000771}

\bibitem{fang2023boron}
M.~Fang, J.~Lacy, A.~Athanasiades, A.~Di~Fulvio, Boron coated straw-based
  neutron multiplicity counter for neutron interrogation of triso fueled
  pebbles, Annals of Nuclear Energy 187 (2023) 109794.

\bibitem{fang2023feasibility}
M.~Fang, A.~Di~Fulvio, Feasibility of neutron coincidence counting for spent
  triso fuel, Annals of Nuclear Energy 193 (2023) 110062.

\bibitem{varshalovich1988quantum}
V.~K. Khersonskii, A.~N. Moskalev, D.~A. Varshalovich, Quantum theory of
  angular momentum, World Scientific, 1988.

\bibitem{weber2003essential}
H.~J. Weber, G.~B. Arfken, Essential mathematical methods for physicists, ISE,
  Elsevier, 2003.

\end{thebibliography}

\appendix

\section{Proof of Rotation-Invariance of Descriptor $\mathcal{H}$}\label{apdx:rotation_invariance}
First we consider the transformation $\mathcal{F}$ that maps a point cloud $\pcx$ to a spherical function $f$ in Eq.~\eqref{eq:sph_fuc_def} and we show that $\mathcal{F}$ is rotation-equivariant, i.e,
    \begin{gather}
        \forall \rotation\in \SO(3), \pcx\in \mathbb{R}^{n\times 3}, \mathcal{F}(\rotation(\pcx))=\rotation(\mathcal{F}(\pcx))
    \end{gather}
\begin{proof}
    Let $g = \mathcal{F}(\rotation(\pcx))$, $h=\rotation(\mathcal{F}(\pcx))$. By definition, $\forall \vecx \in \SPH^2$,
    \begin{gather}
        g(\vecx)=\sum_{i=1}^n \exp(-d^2(\vecx,\rotation\vecx_i)/2\sigma^2),\\
        h(\vecx)=f(\rotation^{-1}\vecx) = \sum_{i=1}^n \exp(-d^2(\rotation^{-1}\vecx,\vecx_i)/2\sigma^2)
    \end{gather}
    We will show $\forall i,d(\vecx,\rotation\vecx_i)=d(\rotation^{-1}\vecx,\vecx_i)$ $\Leftrightarrow \forall i,\|\vecx-\rotation\vecx_i\| = \|\rotation^{-1}\vecx-\vecx_i\|$ $\Leftrightarrow \forall i,\langle\vecx, \rotation\vecx_i\rangle = \langle \rotation^{-1}\vecx, \vecx_i\rangle$.
    Since $\rotation\in \SO(3)$, we have $\rotation^{-1}=\rotation^T$. Therefore, 
    \begin{equation}
        RHS = \langle \rotation^{-1}\vecx, \vecx_i\rangle=\langle \rotation^T\vecx, \vecx_i\rangle = \langle \vecx, \rotation\vecx_i\rangle = LHS
    \end{equation}
\end{proof}

Then we consider the transformation $\mathcal{G}$ that maps a spherical function $f(\vecx)$ to a feature $h$ in Eq.\eqref{eq:feature_definition} and we will show that $\mathcal{G}$ is rotation-invariant, i.e.,
\begin{equation}
    \forall \rotation \in \SO(3), f\in C(S^{2}), \mathcal{G}(\rotation(f)) = \mathcal{G}(f)
\end{equation}
\begin{proof}
    Consider a rotation $\rotation\in \SO(3)$ that maps the unit vector $\vecx$ to $\vecx'$. By definition, the spherical harmonics expansion coefficient in the new coordinate system
    \begin{equation}
        a'_{lm}=\int_{\SPH^2}f(\vecx)Y_{lm}^*(\vecx')d\Omega' =\int_{\SPH^2}f(\vecx)Y_{lm}^*(\rotation\vecx)d\Omega    
    \end{equation}
    $Y_{lm}(\rotation\vecx)$ is related to $Y_{lm}(\vecx)$ through the Wigner D-matrix~\cite{varshalovich1988quantum},
    \begin{equation}
        Y_{lm}(\rotation\vecx)=\sum_{m'=-l}^l [D_{mm'}^l(\rotation)]^* Y_{lm'}(\vecx)
    \end{equation}
    Therefore,
    \begin{equation}
        a'_{lm}=\sum_{m'=-l}^l D_{mm'}^l(\rotation) a_{lm'}
    \end{equation}
    \begin{equation}
    \begin{aligned}
        \sum_{m=-l}^l|a'_{lm}|^2&=\sum_{m=-l}^l (\sum_{m'=-l}^l D_{mm'}^l(\rotation) a_{lm'}) (\sum_{m"=-l}^l D_{mm"}^l(\rotation) a_{lm"})^*\\
        &=\sum_{m'=-l}^l \sum_{m"=-l}^l (\sum_{m=-l}^lD_{mm'}^l(\rotation) [D_{mm"}^l(\rotation)]^*) a_{lm'}a_{lm"}^*
    \end{aligned}
    \end{equation}
    Using the orthogonality of $D_{mm'}^l(\rotation)$~\cite{varshalovich1988quantum}:
    \begin{equation}
            \sum_{m=-l}^l D_{mm'}^l(R) [D_{mm"}^l(\rotation)]^* = \delta_{m'm"}
    \end{equation}
    \begin{equation}
    \begin{aligned}
        &\sum_{m=-l}^l|a'_{lm}|^2=\sum_{m'=-l}^l \sum_{m"=-l}^l \delta_{m'm"} a_{lm'}a_{lm"}^*=\sum_{m'=-l}^l a_{lm'}a_{lm'}^*\\
        =&\sum_{m=-l}^l|a_{lm}|^2
    \end{aligned}
    \end{equation}
    Therefore, for any $l$, we have $\|(\rotation(f))_l\| =\|f_l\|\Rightarrow \mathcal{G}(\rotation(f))=\mathcal{G}(f)$.
\end{proof}

Finally, we show that the descriptor $\mathcal{H}$, which is the combination of $\mathcal{G}$ and $\mathcal{F}$, is rotation-invariant.
\begin{proof}
    \begin{equation}
    \begin{aligned}
        &\forall \rotation\in \SO(3), \pcx\in \mathbb{R}^{n\times 3}, \\
        &\mathcal{H}(\rotation(\pcx))= \mathcal{G}(\mathcal{F}(\rotation(\pcx)))=\mathcal{G}(\rotation(\mathcal{F}(\pcx)))=\mathcal{G}(\mathcal{F}(\pcx))=\mathcal{H}(\pcx)
    \end{aligned}
    \end{equation}
\end{proof}

\section{Proof of Reflection-Invariance of Descriptor $\mathcal{H}$}\label{apdx:reflection_invariance}
Due to the rotation-invariance of $\mathcal{H}$, we will only need to show this for a specific reflection. 

\begin{proof}
Set the reflection plane to be the $x-y$ plane. The spherical function after reflection $f'(\theta,\phi)$ is related to the original spherical function $f(\theta,\phi)$ through:
\begin{equation}
    f'(\theta,\phi)=f(\pi-\theta,\phi)
\end{equation}
The spherical harmonics expansion coefficient of $f'(\theta,\phi)$ is
\begin{equation}
\begin{aligned}
    a'_{lm}&=\langle Y_{lm}(\theta,\phi), f'(\theta,\phi)\rangle =\langle Y_{lm}(\theta,\phi), f(\pi-\theta,\phi)\rangle\\
    &=\langle Y_{lm}(\pi-\theta,\phi), f(\theta,\phi)\rangle
\end{aligned}
\end{equation}
Using the parity of $Y_{lm}$~\cite{weber2003essential}:
\begin{equation}
    Y_{lm}(\pi-\theta,\phi)=(-1)^{l+m}Y_{lm}(\theta,\phi)
\end{equation}
we obtain
\begin{equation}
    a'_{lm}=(-1)^{l+m}a_{lm}\Rightarrow |a'_{lm}|=|a_{lm}|
\end{equation}
Therefore the extracted features are invariant under reflections.
    
\end{proof}

\end{document}